\documentclass[aps,twocolumn,prx,preprintnumbers,showpacs,amsmath,amssymb,superscriptaddress,longbibliography]{revtex4-1}
\usepackage[pdftex, colorlinks, citecolor=blue]{hyperref}   
\usepackage{graphicx}
\usepackage{dcolumn}
\usepackage{threeparttable}
\usepackage{multirow}
\usepackage{booktabs}
\usepackage{txfonts}
\usepackage{xcolor}
\usepackage{bm}
\usepackage{amssymb}
\usepackage{amsmath}
\usepackage{latexsym}
\usepackage{epsfig}
\usepackage{amsbsy}
\usepackage{array}
\usepackage{tabularx}
\usepackage{esvect} 
\usepackage{extarrows}
\usepackage{float}
\usepackage[british]{babel}

\begin{document}

\title{Atomistic mechanisms of phase transitions in all-temperature barocaloric material KPF$_6$}

\author{Jiantao Wang}
\affiliation{Shenyang National Laboratory for Materials Science, Institute of Metal Research, Chinese Academy of Sciences, 110016 Shenyang, China}
\affiliation{School of Materials Science and Engineering, University of Science and Technology of China, 110016 Shenyang, China}

\author{Yi-Chi Zhang}
\affiliation{Shenyang National Laboratory for Materials Science, Institute of Metal Research, Chinese Academy of Sciences, 110016 Shenyang, China}
\affiliation{School of Materials Science and Engineering, University of Science and Technology of China, 110016 Shenyang, China}

\author{Yan Liu}
\affiliation{Shenyang National Laboratory for Materials Science, Institute of Metal Research, Chinese Academy of Sciences, 110016 Shenyang, China}
\affiliation{School of Materials Science and Engineering, University of Science and Technology of China, 110016 Shenyang, China}

\author{Hongkun Deng}
\affiliation{Shenyang National Laboratory for Materials Science, Institute of Metal Research, Chinese Academy of Sciences, 110016 Shenyang, China}
\affiliation{School of Materials Science and Engineering, University of Science and Technology of China, 110016 Shenyang, China}

\author{Mingfeng Liu}
\affiliation{Shenyang National Laboratory for Materials Science, Institute of Metal Research, Chinese Academy of Sciences, 110016 Shenyang, China}

\author{Yan Sun}
\affiliation{Shenyang National Laboratory for Materials Science, Institute of Metal Research, Chinese Academy of Sciences, 110016 Shenyang, China}

\author{Bing Li}
\affiliation{Shenyang National Laboratory for Materials Science, Institute of Metal Research, Chinese Academy of Sciences, 110016 Shenyang, China}

\author{Xing-Qiu Chen}
\email{xingqiu.chen@imr.ac.cn}
\affiliation{Shenyang National Laboratory for Materials Science, Institute of Metal Research, Chinese Academy of Sciences, 110016 Shenyang, China}

\author{Peitao Liu}
\email{ptliu@imr.ac.cn}
\affiliation{Shenyang National Laboratory for Materials Science, Institute of Metal Research, Chinese Academy of Sciences, 110016 Shenyang, China}

\begin{abstract}
Conventional barocaloric materials typically exhibit limited operating temperature ranges.
In contrast, KPF$_6$ has recently been reported to achieve an exceptional all-temperature barocaloric effect (BCE) via pressure-driven phase transitions.
Here, we elucidate the atomistic mechanisms underlying the phase transitions through first-principles calculations and machine-learning potential accelerated molecular dynamics simulations.
We identify four distinct phases:
the room-temperature cubic (C) plastic crystal characterized by strong fluorine orientational disorder (FOD) and anharmonicity,
the intermediate-temperature monoclinic (M-II) phase with decreasing FOD,
the low-temperature monoclinic (M-I) phase with suppressed FOD,
and the fully ordered rhombohedral (R) phase under pressure.
Phonon calculations confirm the dynamic stability of the M-II, M-I, and R phases at 0 K, whereas the C phase requires thermal fluctuations for stabilization.
Under pressure, all the C, M-II, and M-I phases transform to the R phase,
which are driven by cooperative PF$_6$ octahedral rotations coupled with lattice modulations.
These pressure-induced phase transitions result in persistent isothermal entropy changes across a wide temperature range,
thereby explaining the experimentally observed all-temperature BCE in this material.
Hybrid functional calculations reveal wide-bandgap insulating behavior across all phases.
This work deciphers the interplay between FOD, anharmonicity, and phase transitions in KPF$_6$,
providing important insights for the design of BCE materials with broad operational temperature spans.
\end{abstract}

\maketitle

\section{Introduction}

Caloric effects, driven by external fields such as magnetic fields~\cite{Shen2009},
electric fields~\cite{Scott2011}, uniaxial stress~\cite{Mechanocaloric2017}, or hydrostatic pressure~\cite{barocaloric2010},
present a promising avenue for solid-state refrigeration, offering a sustainable alternative to conventional vapor-compression systems.
Among these, the barocaloric effect (BCE)---characterized by isothermal entropy changes or adiabatic temperature changes
upon pressure application or removal---stands out due to its universality, scalability, and large thermal responses~\cite{barocaloric2010,Review_Cirillo2022}.
To date, the BCE has been demonstrated in various systems such as shape memory alloys~\cite{barocaloric2010,PhysRevMaterials.3.044406,Stern-Taulats2015,Review_Cirillo2022},
superionic compounds~\cite{Cazorla2016,AgI_NC2017,PhysRevMaterials.4.015403},
and organic-inorganic hybrid perovskites~\cite{hybridperovskite_NC2017,JPCL2017,Salgado-Beceiro2020,LiAFM2021}.
In recent years, plastic crystals have received growing interest since the discovery of colossal BCEs in
these materials~\cite{Li_Nature2019,Lloveras2019,Aznar2020,Wanghui_NC2020,Sau_ScientificReports2021,LibingNC2022,
Salvatori2022,ZhangAFM2022,LibingSA2023,EscorihuelaSayalero2024,PiperScience2025}.
These plastic crystals exhibit a unique structural duality: their molecular or ionic subunits display orientational disorder
while maintaining long-range positional order in their center of mass.
The orientational disorder-order phase transitions in plastic crystals combined with their high compressibility and strong anharmonicity
enable giant isothermal entropy changes, positioning them ideal candidates for BCE-driven refrigeration~\cite{Li_Nature2019}.

However, most known caloric materials exhibit a relatively narrow operating temperature span~\cite{LibingNC2025}.
For example, the prototype barocaloric material neopentylglycol exhibits only a modest increase of 10 K in the phase transition temperature
from the cubic phase to the monoclinic phase when subjected to a pressure of 100 MPa~\cite{LibingNC2025}.
This limitation---where materials are functional only within a restricted temperature range---necessitates multi-stage configurations in caloric refrigerators to achieve broader cooling spans,
leading to increased system complexity and costs~\cite{LibingNC2025}.
Recently, this challenge has been overcome by the discovery of an all-temperature BCE in KPF$_6$~\cite{LibingNC2025}.
Remarkably, KPF$_6$ demonstrates an exceptionally wide BCE temperature span from 77 K to 300 K~\cite{LibingNC2025}.
The all-temperature BCE is linked to persistent pressure-induced phase transitions, as corroborated by pressure-dependent neutron powder diffraction (NPD),
Raman scattering analyses, and first-principles calculations~\cite{LibingNC2025}.

\begin{figure*}
\begin{center}
\includegraphics[width=0.85\textwidth,trim = {0.0cm 0.0cm 0.0cm 0.0cm}, clip]{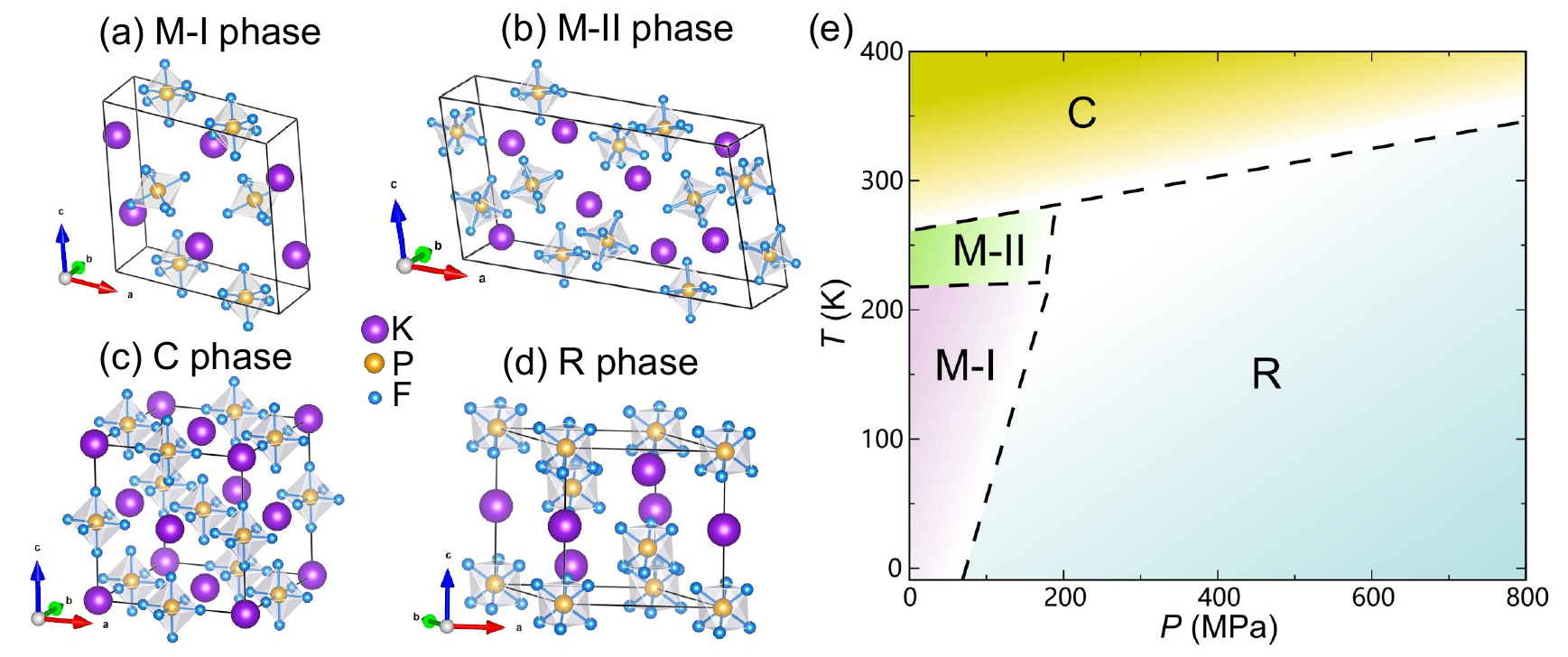}
\end{center}
\caption{(a)-(d) Crystal structures of the four phases of KPF$_6$.
(e) The experimentally-determined temperature-pressure phase diagram of KPF$_6$, adapted from Ref.~\cite{LibingNC2025}.
}
\label{Fig1_str_diagram}
\end{figure*}

While the temperature-pressure phase diagram for KPF$_6$ has been experimentally established
through pressure-dependent Raman spectroscopy and neutron diffraction~\cite{Zhang2023,LibingNC2025},
the atomistic mechanisms underlying the phase transitions across its
four phases (see Fig.~\ref{Fig1_str_diagram})---room-temperature face-centered cubic (C), intermediate-temperature monoclinic (M-II), low-temperature monoclinic (M-I),
and high-pressure rhombohedral (R)---remain unresolved.
Additionally, thorough investigations into the dynamic, thermodynamic, and electronic properties of these phases are lacking.

To address these gaps, we developed a robust machine-learning potential (MLP) for KPF$_6$ that can accurately describe all four phases across wide temperature and pressure ranges.
This MLP enables accurate and efficient identification of crystal structures, exploration of dynamic and thermodynamic properties, and revelation of phase transitions under pressure.
Phonon calculations confirm the dynamic stability of the M-II, M-I, and R phases at 0 K, with distinct flat phonon bands across broad frequency ranges arising mainly from PF$_6$ vibrations.
In contrast, the thermally-disordered C phase displays pronounced anharmonicity and achieves dynamic stability only at elevated temperatures.
Interestingly, the fluorine atoms within the PF$_6$ octahedra exhibit significant orientational disorder in the C phase.
This disorder diminishes as the system transitions to the M-II phase at lower temperatures, and becomes suppressed in the M-I phase at 50 K.
Under pressure, all the C, M-II, and M-I phases transform to the fully ordered R phase,
leading to persistent isothermal entropy changes across a wide temperature range, which are responsible for the all-temperature BCE in this material.
The phase transitions proceed through the cooperative rotation of the PF$_6$ octahedra accompanied by the lattice modulations.
Finally, hybrid functional calculations of electronic band structures demonstrate wide-bandgap insulating behavior across all phases.

\section{Computational details}

\subsection{First-principles calculations}

First-principles density functional theory (DFT) calculations were conducted using the Vienna ab initio simulation package (VASP)~\cite{PhysRevB.54.11169}.
The interactions between nuclei and valence electrons were described by the projector augmented wave pseudopotentials recommended by VASP~\cite{PhysRevB.50.17953,PhysRevB.59.1758}.
We employed the PBEsol exchange-correlation functional~\cite{PhysRevLett.100.136406},
which typically provides an improved description of the equilibrium properties of solids~\cite{PhysRevLett.100.136406}.
To ensure convergence of the total energy to better than 1 meV/atom, a high plane wave energy cutoff of 700 eV was utilized.
A $\Gamma$-centered $k$-point grid with a reciprocal-space resolution of 0.21 \AA$^{-1}$ was used for sampling the Brillouin zone.
The convergence criteria for total energy and ionic forces were set to $10^{-6}$ eV and 5 meV/$\AA$, respectively.
The Gaussian smearing method with a smearing width of 0.05 eV was applied.
The electronic band structures were computed using the hybrid functional HSE06~\cite{HSE06}.
Phonon dispersions and density of states were calculated using finite displacements with the Phonopy code~\cite{Togo2015-pho}.
To determine the energy barrier for phase transitions, we employed the variable cell nudged elastic band (VCNEB) method~\cite{QIAN20132111}.

\subsection{Machine-learning potential development}

The development of machine-learning potential followed the scheme outlined in our previous works~\cite{PhysRevLett.130.078001,Liu2024,PhysRevLett.134.178001}.
Specifically, the training set generation consisted of two processes.
The first process employed the on-the-fly active learning procedure as implemented in the VASP~\cite{JinnouchiPRL2019,JinnouchiPRB2019}.
This method is based on the kernel-based Bayesian regression and allows to automatically select the representative structures
during the \emph{ab initio} molecular dynamics (AIMD) simulations~\cite{JinnouchiPRL2019,JinnouchiPRB2019}.
The cutoff radii were set to 6 $\AA$ and 5 $\AA$ for the two-body and three-body descriptors, respectively.
The number of radial basis functions was set to 10 for both the two-body and three-body descriptors.
The width of the Gaussian functions used for broadening the atomic distributions of the descriptors was set to 0.5 $\AA$.
The on-the-fly AIMD samplings were performed by heating the four phases of KPF$_6$ from 50 K to 600 K under pressures of 0, 0.5, and 1.0 GPa.
The initial supercell structures were built with sixteen formula units based on the DFT relaxed structures. A time step of 2 fs was employed.
The isothermal-isobaric ensemble was employed using a Langevin thermostat~\cite{allenThermostatMolecularDynamics2007}
combined with the Parrinello-Rahman method~\cite{parrinelloCrystalStructurePair1980,parrinelloPolymorphicTransitionsSingle1981}.

The second process involved multiple active learning cycles utilizing the fitted moment tensor potential (MTP)~\cite{Alexander2016},
which is based on Shapeev's generalized D-optimality criterion~\cite{mtpal1}.
This approach leverages the high inference speed of the linearly parameterized MTP compared to kernel-based methods~\cite{zuoPerformanceCostAssessment2020}.
Additionally, the D-optimality criterion facilitates the efficient assessment of whether an unseen structure falls within the reliable interpolation regime
or the risky extrapolation regime, eliminating the need for expensive DFT calculations \emph{a priori}~\cite{mtpal1}.
The active learning cycles started by generating initial structures that encompass a broad range of diversity,
including the pristine four phases and the interpolated intermediate structures among them.
Subsequently, a series of 200-ps MD simulations were performed at various temperatures and pressures using the LAMMPS code~\cite{thompsonLAMMPSFlexibleSimulation2022},
interfaced with the MTP~\cite{Novikov_2021}. The Langevin thermostat~\cite{allenThermostatMolecularDynamics2007} and the Parrinello-Rahman method~\cite{parrinelloCrystalStructurePair1980,parrinelloPolymorphicTransitionsSingle1981}
were employed to control the temperature and pressure of the system, with a time step set to 2 fs.
Configurations with an extrapolation grade exceeding 10 were identified,
and the energies, forces, and stress tensors of these selected configurations were computed using PBEsol.
These results were incorporated into the training set at the end of each active learning cycle, followed by a refitting of the MTP model.
This iterative process continued until no configurations reached the specified threshold for extrapolation grade.
To enhance efficiency, the MTP was fitted with optimized basis sets by refining the contraction process of moment tensors using our in-house code (IMR-MLP)~\cite{wang2024}.
This refinement significantly accelerates calculations without sacrificing accuracy~\cite{wang2024}.
When constructing the MTP basis functions, a cutoff radius of 6.0~\AA~was employed, with the number of radial basis functions set to 8.
The maximum number of moment tensors involved in contraction was limited to 4, allowing for the consideration of up to five-body interactions.
Through these multiple active learning cycles, the sampled phase space was significantly enhanced, resulting in improved transferability of the MTP.

\begin{figure}
\begin{center}
\includegraphics[width=0.48\textwidth,trim = {0.0cm 0.0cm 0.0cm 0.0cm}, clip]{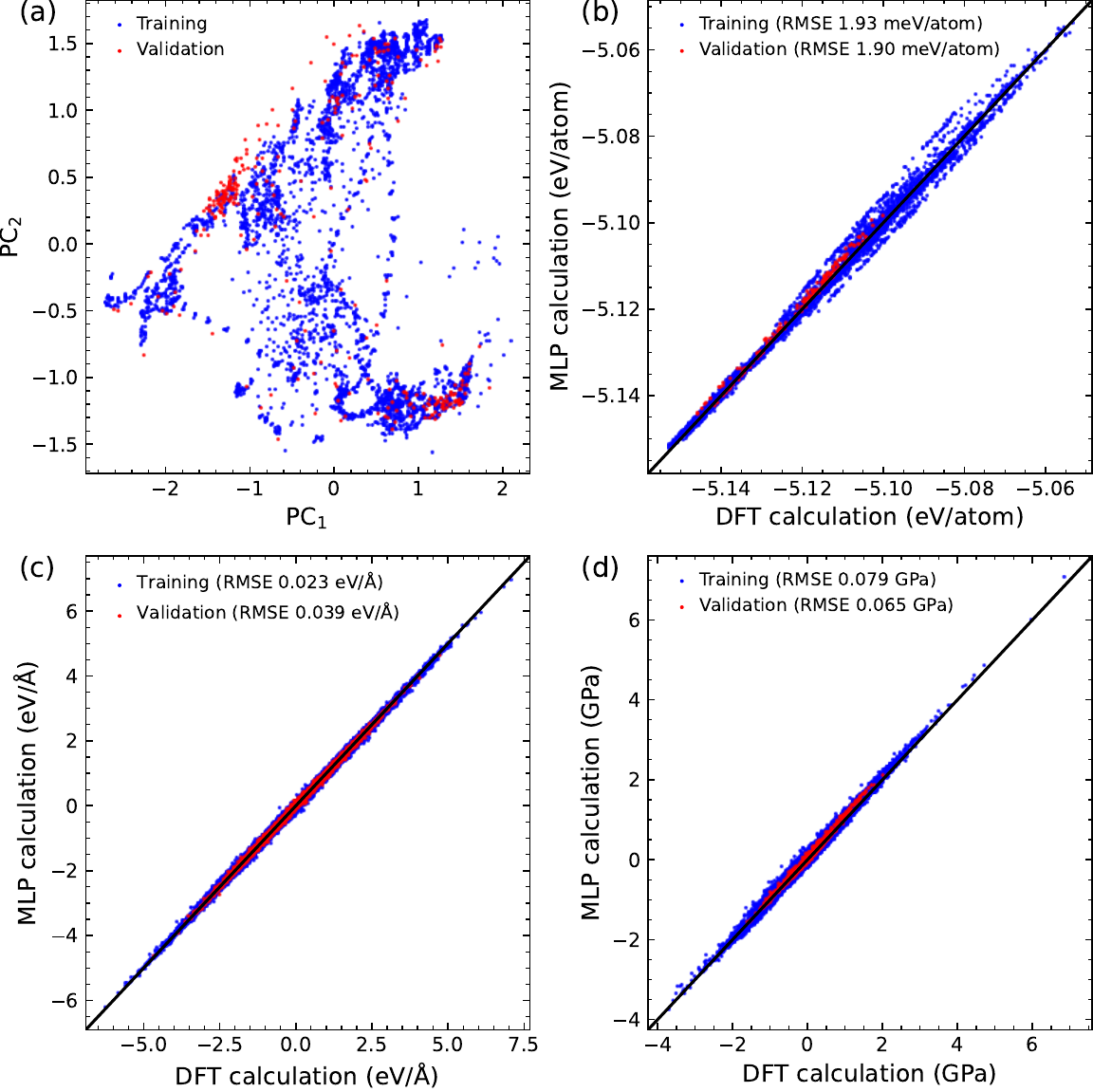}
\end{center}
		\caption{(a) The kernel principal component analysis of training and validation structures.
			(b)-(d) MLP predicted energies, forces, and stress tensors against DFT results.
The training and validation datasets are indicated by the blue and red points, respectively.
}
\label{Fig2_MLP_validation}
\end{figure}

The eventual training dataset includes 3418 structures.
Additionally, a validation dataset consisting of 320 structures was constructed.
These validation structures were randomly sampled from MD simulations,
as well as from structural perturbations applied to interpolated configurations between different phases.
The training and validation structures were analyzed by principal component analysis (PCA) using the smooth overlap of atomic positions as local structure descriptors~\cite{chengMappingMaterialsMolecules2020}.
The kernel PCA analysis, as shown in Fig.~\ref{Fig2_MLP_validation}(a), demonstrates that the validation dataset covers a fraction of the training set.
The developed MTP exhibits a high accuracy, with validation root-mean-square errors
for energies, forces, and stress tensors being 1.90 meV/atom, 0.039 eV/$\AA$, and 0.065 GPa, respectively [see Figs.~\ref{Fig2_MLP_validation}(b)-(d)].

\subsection{Entropy calculations}

Considering that the vibrational-orientational couplings are normally negligible,
the total entropy $S$ at pressure $p$ and temperature $T$ can be expressed as~\cite{EscorihuelaSayalero2024}
\begin{equation}
	\label{eq:Stot}
S(p,T) = S_\text{vib}(p,T) + S_\text{ori}(p,T),
\end{equation}
where $S_\text{vib}$ and $S_\text{ori}$  represent the vibrational and orientational contributions, respectively.

The vibrational entropy $S_\text{vib}$ is computed by~\cite{EscorihuelaSayalero2024}
\begin{equation}
	\begin{split}
		S_\text{vib}(p,T) = & - 3 N_\text{at} k_B \int_0^\infty \ln \left[ 2\sinh \left(\frac{\hbar \omega}{2k_B T}\right) \right] \rho(\omega,p) \mathrm{d}\omega + \\
		& 3N_\text{at} \int_0^\infty \frac{\hbar \omega}{2T}\tanh^{-1}\left(\frac{\hbar\omega}{2k_BT}\right)\rho(\omega,p)\mathrm{d}\omega.
	\end{split}
\end{equation}
Here, $N_\text{at}$ is the number of atoms in the simulation cell,
$k_B$ is the Boltzmann's constant,
$\hbar$ is the reduced Planck constant,
and $\rho(\omega,p)$ represents the vibrational density of states,
which is estimated as the Fourier transform of the velocity autocorrelation function $C_v(t,p)$
\begin{equation}
	\rho (\omega,p) = \int_{0}^{\infty} C_v(t,p) e^{i\omega t} \mathrm{d}t.
\end{equation}
The $C_v(t,p)$ was calculated using velocity data obtained from a 20 ps MD trajectory in the \emph{NpT} ensemble, sampled at intervals of 5 fs.

The orientational entropy $S_\text{ori}$ is calculated as follows.
We first compute the quaternions $q = \{w,x,y,z\}$ representing the rotation of each PF$_6$ octahedron
and restrict the domain to $w > 0$ since $q$ and $-q$ represent the same rotation.
Assuming each PF$_6$ octahedron rotates independently, the orientational entropy per PF$_6$ octahedron can be calculated as
\begin{equation}
	\label{eq:Sori}
	S_\text{ori} = -k_B \int_{\mathbb{S}^3_+} P(q) \ln P(q)  \mathrm{d}q,
\end{equation}
where $P(q)$ is the probability distribution function of the rotation quaternions,
and the integration is performed over the positive hemisphere of the 3-sphere, denoted as $\mathbb{S}^3_+$.
In practical calculations, we constructed a uniform Fibonacci grid over $\mathbb{S}^3_+$
using the algorithm described in Ref.~\cite{10.1109/CVPR52688.2022.00811}.
The discrete form of Eq.~\eqref{eq:Sori} can thus be derived as
\begin{equation}
	S_\text{ori} = -k_B \sum_{i=1}^{N_\text{bins}} \frac{n_i}{N} \ln \frac{n_i}{N} + k_B \ln \left( \frac{\pi^2}{N_\text{bins}} \right),
\end{equation}
where $N$ represents the total number of quaternion samples,
$N_\text{bins}$ is the number of histogram bins,
and $n_i$ denotes the count of samples in bin $i$.
In practice, 4000 grid points were used to sample the quaternion space.
Orientational data were sampled from a 5 ns \emph{NpT}-MD trajectory
using the supercells consisting of 2306 atoms for the M-I, M-II, and R phases, and 2048 atoms for the C phase.

\section{Results and discussion}

\subsection{Crystal structures}\label{sec:crystal_str}

\begin{table}
\caption{The lattice parameters of  the four phases of KPF$_6$ predicted by PBEsol at 0 K and MD simulations at 300 K.
The experimental data are given for comparison.
It is important to note the experimental conditions under which the data were obtained for each phase: the M-I phase data were collected at 3.5 K and 0 GPa~\cite{LibingNC2025},
the M-II phase data at 250 K and 0 GPa~\cite{LibingNC2025}, the C phase data at 300 K and 0 GPa~\cite{Zhang2023}, and the R phase data at 10 K and 0.4 GPa~\cite{LibingNC2025}.
}
\begin{ruledtabular}
\begin{tabular}{clcccc}
   & & $a$  (\AA)  & $b$ (\AA)  &$c$ (\AA) & $\beta$/$\gamma$ ($^{\rm o}$)  \\
\hline
\multirow{2}{1.5cm}{M-I ($C2/c$)}
& Cal. (0 K, 0 GPa) &9.655 & 5.106 & 9.617  & 104.00    \\
& Exp. (3.5 K, 0 GPa)\cite{LibingNC2025} & 9.417 & 4.909 &  9.459  & 103.40    \\
\hline
\multirow{2}{1.5cm}{M-II ($Cc$)}
& Cal. (0 K, 0 GPa) & 18.121 & 5.306 & 9.626  & 102.07    \\
& Exp. (250 K, 0 GPa)\cite{LibingNC2025} & 18.130 & 5.384 & 9.583  & 101.08    \\
\hline
\multirow{3}{1.5cm}{C ($Fm\bar{3}m$)}
& Cal. (0 K, 0 GPa) & 8.543 & 8.543 & 8.543  & 90.00    \\
& Cal. (300 K, 0 GPa, MD) & 8.071 & 8.071 & 8.071  & 90.00    \\
& Exp. (300 K, 0 GPa)\cite{Zhang2023} & 7.786 & 7.786 & 7.786  & 90.00    \\
\hline
\multirow{3}{1.5cm}{R ($R\bar{3}$)}
& Cal. (0 K, 0 GPa) & 7.317 & 7.317 & 7.094  & 120.00    \\
& Cal. (0 K, 0.4 GPa) & 7.226 & 7.226 & 7.052  & 120.00    \\
& Exp. (10 K, 0.4 GPa)\cite{LibingNC2025} & 7.096 & 7.096 &  7.005  & 120.00    \\
\end{tabular}
\end{ruledtabular}
\label{Table1}
\end{table}

Figure~\ref{Fig1_str_diagram} depicts the crystal structures for the M-I, M-II, and R phases of KPF$_6$ and the experimentally-determined temperature-pressure phase diagram.
The M-I phase is a low-temperature phase exhibiting a monoclinic crystal structure with the space group $C2/c$.
The M-II phase represents an intermediate-temperature phase that also exhibits a monoclinic crystal structure, albeit with reduced symmetry (space group $Cc$).
The C phase is the room-temperature phase, which possesses a face-centered cubic rock-salt structure.
Lastly, the R phase is a high-pressure phase featuring a rhombohedral crystal structure with the space group $R\bar{3}$.
The simulated X-ray diffraction (XRD) patterns of the four phases are presented in Supplementary Information Fig.~S1~\cite{SM}.

We note that the crystal structures of the C and R phases, including their symmetry and atomic positions,
have been previously determined through temperature-dependent single-crystal XRD and NPD~\cite{Zhang2023,LibingNC2025}.
In contrast, the M-I and M-II phases, which are characterized by lower crystal symmetries and larger unit cells,
pose significant challenges in identifying their space groups and atomic positions~\cite{Huber1997,Zhang2023}.
This challenge has only been addressed recently~\cite{LibingNC2025}
through the application of a MLP-accelerated generic structure search approach~\cite{Glass2006}, combined with experimental constraint information~\cite{LibingNC2025}.
Specifically, the experimental constraints, such as the minimum interatomic distances between different atom types,
significantly reduced the search space, facilitating a more efficient and accurate determination of the complex crystal structures.
The lattice parameters of the four phases of KPF$_6$, predicted using the PBEsol functional, are provided in Table~\ref{Table1}.
One can observe good agreement between the predicted values and the experimental results.
For more detailed information regarding the crystal structures, we refer to the Supplementary Information Tables S1-S4~\cite{SM}.

\begin{figure}
\begin{center}
\includegraphics[width=0.48\textwidth,trim = {0.0cm 0.0cm 0.0cm 0.0cm}, clip]{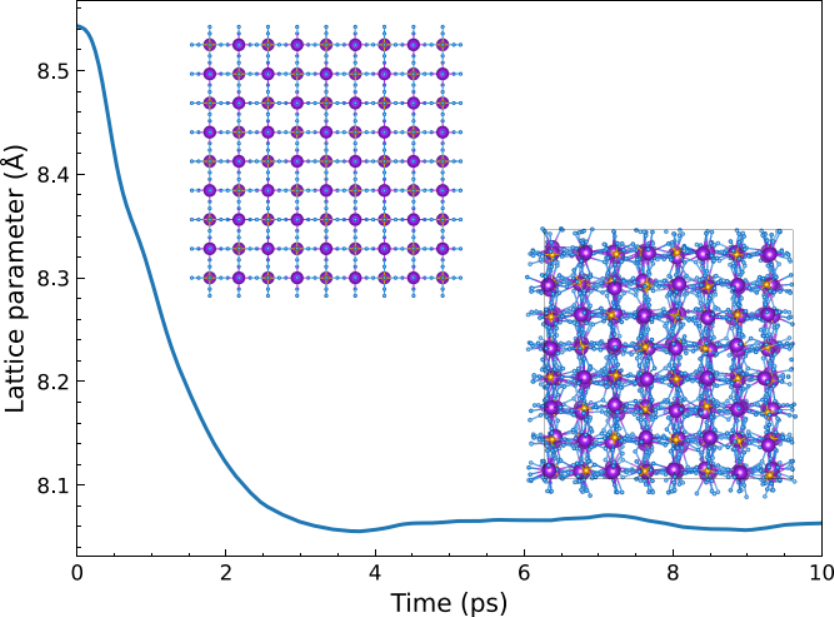}
\end{center}
\caption{Time evolution of the lattice parameter for the C phase during a MD simulation at 300 K and 0 GPa.
Insets show the top view of the initial configuration obtained from PBEsol relaxation and the final equilibrated configuration from the MD simulation,
highlighting the plastic crystal nature of the C phase characterized by significant orientational disorder of the fluorine atoms.
}
\label{Fig3_Lattice_C_NPT}
\end{figure}

It is important to note that the highly symmetric structure of the C phase
depicted in Fig.~\ref{Fig1_str_diagram}(c) is dynamically unstable at 0 K (Supplementary Information Fig.~S2~\cite{SM}).
In fact, this instability is indicative of fluorine orientational disorder (FOD), allowing it to be regarded as an ionic plastic crystal.
The manifestation of diffuse scattering for this phase has been observed in neutron diffraction experiment~\cite{Zhang2023}.
Indeed, when conducting MD simulations at 300 K using the PBEsol-relaxed ordered structure as the initial configuration,
the PF$_6$ octahedra exhibit strong FOD (see Fig.~\ref{Fig3_Lattice_C_NPT}).
This disorder leads to a reduction in lattice parameters, bringing them into closer alignment with the experimental values (see Table~\ref{Table1}).

\subsection{Energetic properties}

\begin{figure}
\begin{center}
\includegraphics[width=0.49\textwidth,trim = {0.0cm 0.0cm 0.0cm 0.0cm}, clip]{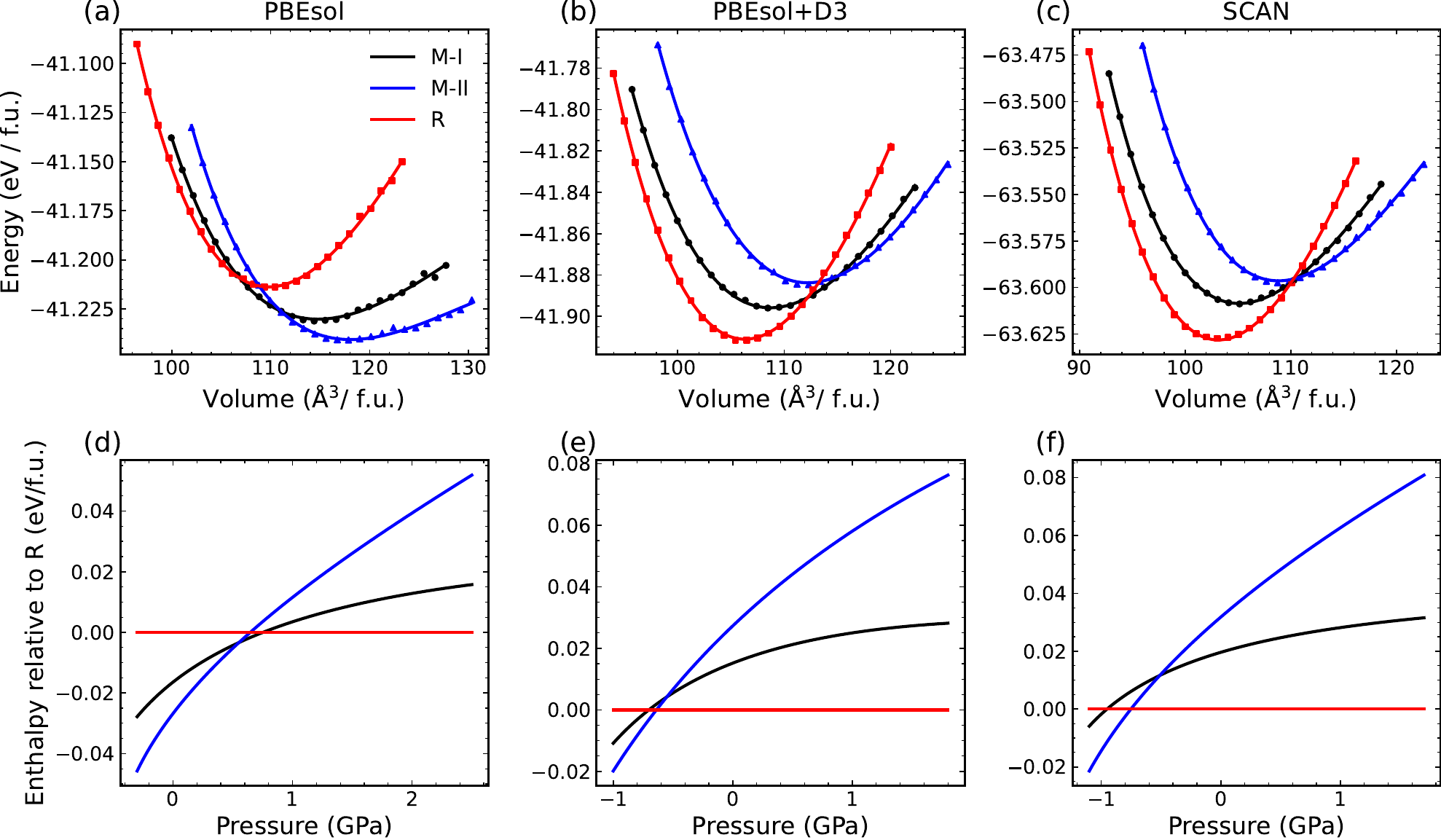}
\end{center}
\caption{Energy-volume curves for the M-I, M-II, and R phases of KPF$_6$, calculated at 0 K using the (a) PBEsol, (b) PBEsol+D3, and (c) SCAN functionals.
(d)-(f) Corresponding enthalpy-pressure curves of the M-I and M-II phases relative to the R phase.
}
\label{Fig4_E_vs_V}
\end{figure}

Figure~\ref{Fig4_E_vs_V}(a) presents the energy-volume curves for the four phases of KPF$_6$ at 0 K predicted by PBEsol.
As expected, the M-I and M-II phases have larger equilibrium volumes than the high-pressure R phase.
The calculated transition pressures from M-II and M-I to the R phase are 0.66 GPa and 0.71 GPa, respectively [Fig.~\ref{Fig4_E_vs_V}(d)].
We note that the C phase exhibits a significantly overestimated equilibrium volume and total energy at 0 K (Supplementary Information Fig.~S3~\cite{SM}), which can be attributed to
the neglect of vibrational and orientational disorder effects of the F atoms, as discussed earlier (Sec.~\ref{sec:crystal_str}).
Notably, PBEsol incorrectly identifies the M-II phase as the ground state, contradicting experimental observations~\cite{LibingNC2025},
though the energy difference between M-I and M-II is as small as 1.3 meV per atom.

Given the presence of FOD in KPF$_6$, we also assessed the role of van der Waals (vdW) interactions.
Including vdW corrections via Grimme's D3 method~\cite{10.1063/1.3382344} restores the correct energetic ordering between the M-I and M-II phases.
However, PBEsol+D3 introduces a new discrepancy: it erroneously predicts the R phase as the ground state [Fig.~\ref{Fig4_E_vs_V}(b)].
This means that a negative pressure is needed to induce the transition from the M-I and M-II phases to the R phase [Fig.~\ref{Fig4_E_vs_V}(e)],
again conflicting with experiments~\cite{LibingNC2025}.
We further examined the strongly constrained appropriately normed (SCAN) functional~\cite{PhysRevLett.115.036402},
which yields energetics similar to PBEsol+D3 [Figs.~\ref{Fig4_E_vs_V}(c) and (f)].
These results suggest that none of the tested exchange-correlation functionals can simultaneously describe all phases accurately.
Since the pressure-driven phase transition is the primary focus of this work, we ultimately adopted the PBEsol functional,
as it provides a more reliable description of the pressure-driven transition behavior despite its limitations in ground-state energetics.

\subsection{Dynamical properties}

\begin{figure}
\begin{center}
\includegraphics[width=0.46\textwidth,trim = {0.0cm 0.0cm 0.0cm 0.0cm}, clip]{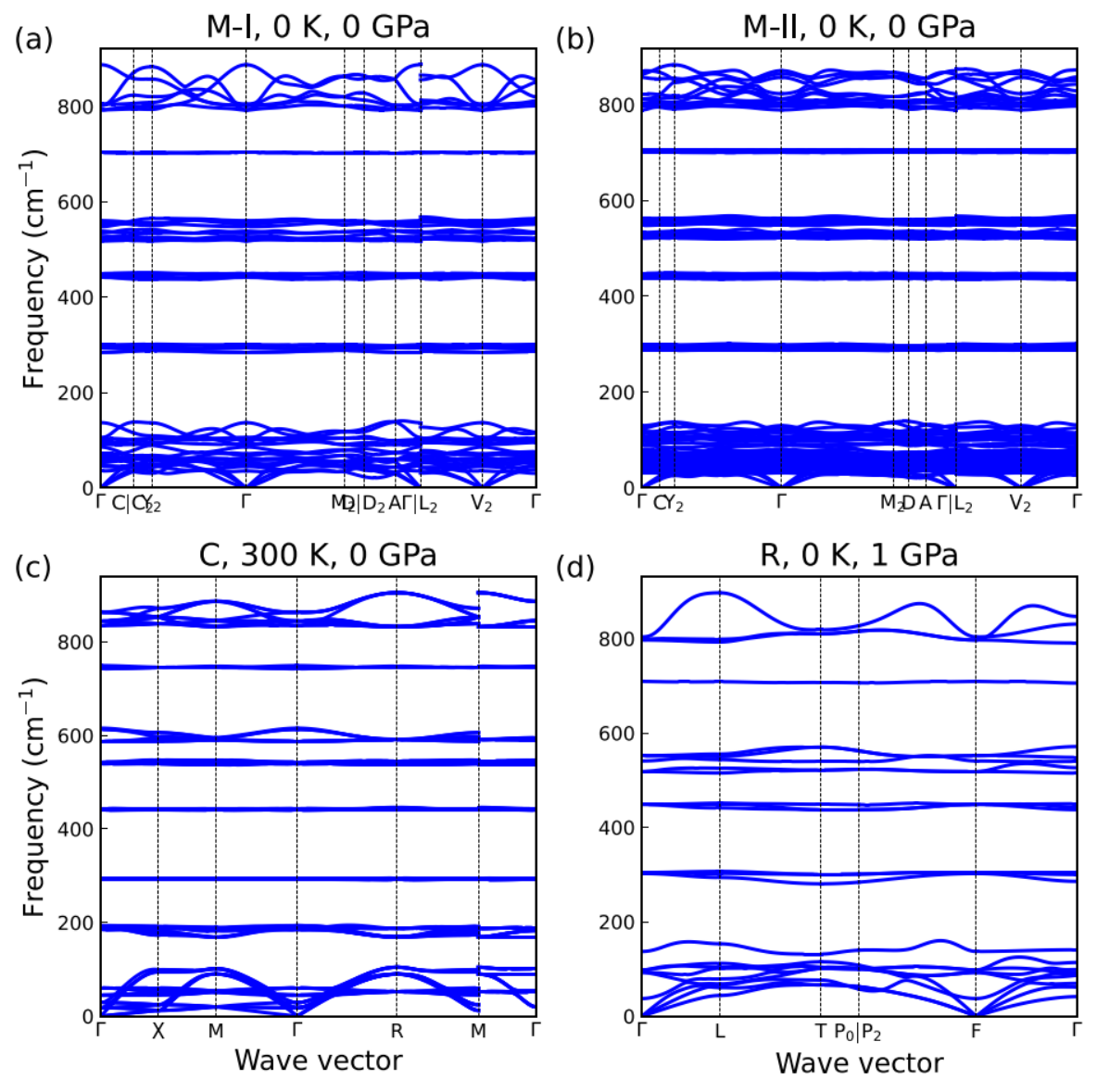}
\end{center}
\caption{Phonon dispersion relationships of the four phases of KPF$_6$ calculated using the MTP.
(a) The M-I phase at 0 K and 0 GPa. (b) The M-II phase at 0 K and 0 GPa. (c) The C phase at 300 K and 0 GPa (obtained by the SSCHA method). (d) The R phase at 0 K and 1 GPa.
}
\label{Fig5_phonon}
\end{figure}

The phonon dispersion relationships of the four phases of KPF$_6$ calculated using the MTP are shown in Fig.~\ref{Fig5_phonon},
exhibiting relatively flat and well-separated bands characteristic of plastic crystals with weak phonon-phonon interactions.
The ground-state M-I phase is dynamically stable at 0 K and 0 GPa [Fig.~\ref{Fig5_phonon}(a)], as expected for the equilibrium structure.
Interestingly, the intermediate-temperature M-II phase also maintains dynamical stability at 0 K [Fig.~\ref{Fig5_phonon}(b)], despite being the intermediate-temperature phase.
In contrast, the high-temperature C phase is dynamically unstable at 0 K and 0 GPa, exhibiting imaginary phonon modes at the harmonic level (Supplementary Information Fig.~S2~\cite{SM}), which reveals its strong anharmonicity. However, when anharmonic phonon-phonon interactions are included at 300 K using the stochastic self-consistent harmonic approximation (SSCHA)~\cite{monacelliStochasticSelfconsistentHarmonic2021}, the C phase becomes dynamically stable, consistent with experimental observations~\cite{LibingNC2025} [Fig.~\ref{Fig5_phonon}(c)]. Finally, the high-pressure R phase demonstrates dynamical stability at both 1 GPa [Fig.~\ref{Fig5_phonon}(d)]
and 0 GPa (Supplementary Information Fig.~S2~\cite{SM}), indicating its robust stability across pressure conditions.

\begin{figure}
\begin{center}
\includegraphics[width=0.45\textwidth,trim = {0.0cm 0.0cm 0.0cm 0.0cm}, clip]{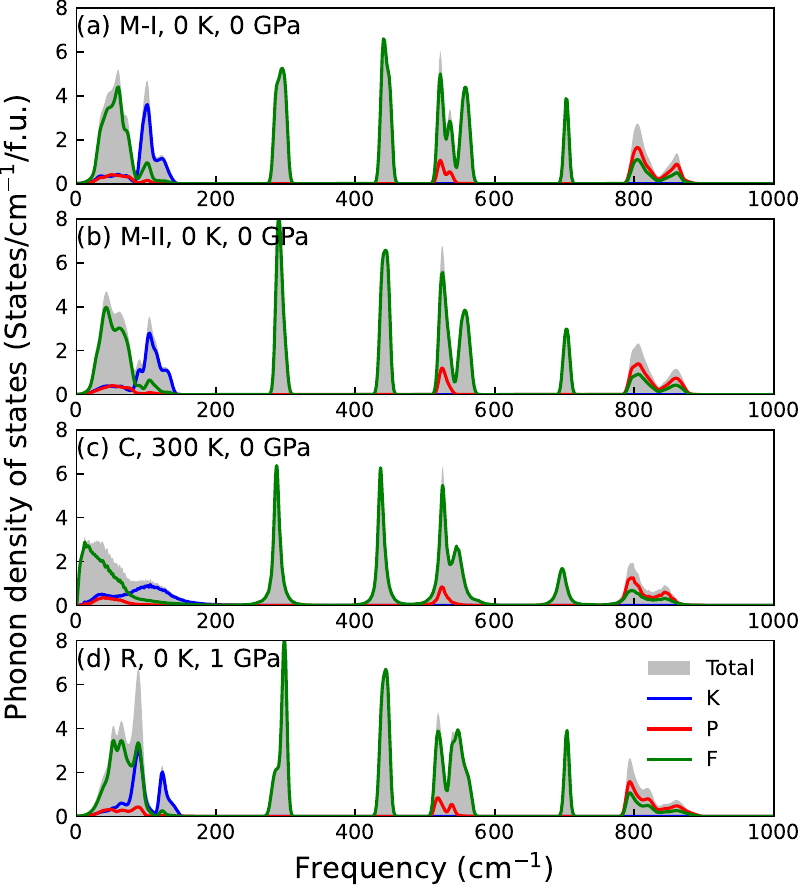}
\end{center}
\caption{Atom-resolved phonon density of states of the four phases of KPF$_6$ calculated using the MTP.
(a) The M-I phase at 0 K and 0 GPa. (b) The M-II phase at 0 K and 0 GPa. (c) The C phase at 300 K and 0 GPa (obtained by MD). (d) The R phase at 0 K and 1 GPa.
}
\label{Fig6_phononDOS}
\end{figure}

We further analyzed the atomic contributions to the phonon modes.
The computed the atom-resolved phonon density of states (DOSs) for the four phases of KPF$_6$ are shown in Fig.~\ref{Fig5_phonon}.
All four phases exhibit similar frequency ranges in their phonon DOSs.
The K atoms primarily vibrate within the 200 cm$^{-1}$ range, while the P atoms cover a relatively broad spectrum of frequencies,
encompassing low, intermediate, and high frequencies.
Notably, the F atoms dominate most frequency regions, with those flat phonon bands exclusively attributed to their vibrations.
Since the four phases exhibit similar atom-resolved DOSs,
we will focus only on the R phase as an example to detail the atomic contributions (Supplementary Information Fig.~S4~\cite{SM}).
The low-frequency phonon modes (below 200 cm$^{-1}$) arise from the rotation of the PF$_6$ octahedra and the vibration of K atoms.
The phonon modes around 300 cm$^{-1}$ and 470 cm$^{-1}$ originate from the bending motions of the PF$_6$ octahedra,
corresponding to the group representations of $F_{2u}$ and $F_{2g}$, respectively.
The phonon modes ranging from 500 to 600 cm$^{-1}$ are associated with the asymmetric stretching and bending of the PF$_6$ octahedra.
The separate strong peak at about 720 cm$^{-1}$  is linked to the $A_{1g}$ mode, which reflects the symmetric stretching vibration of the PF$_6$ octahedra.
The highest phonon modes above 800 cm$^{-1}$ correspond to the stretching of the PF$_6$ octahedra, accompanied by the vibrations of P atoms.
These theoretical phonon mode analyses align well with the experimental Raman spectra results~\cite{LibingNC2025}.

\subsection{Pressure-induced phase transitions and entropy changes}

\begin{figure*}
\begin{center}
\includegraphics[width=0.6\textwidth,trim = {0.0cm 0.0cm 0.0cm 0.0cm}, clip]{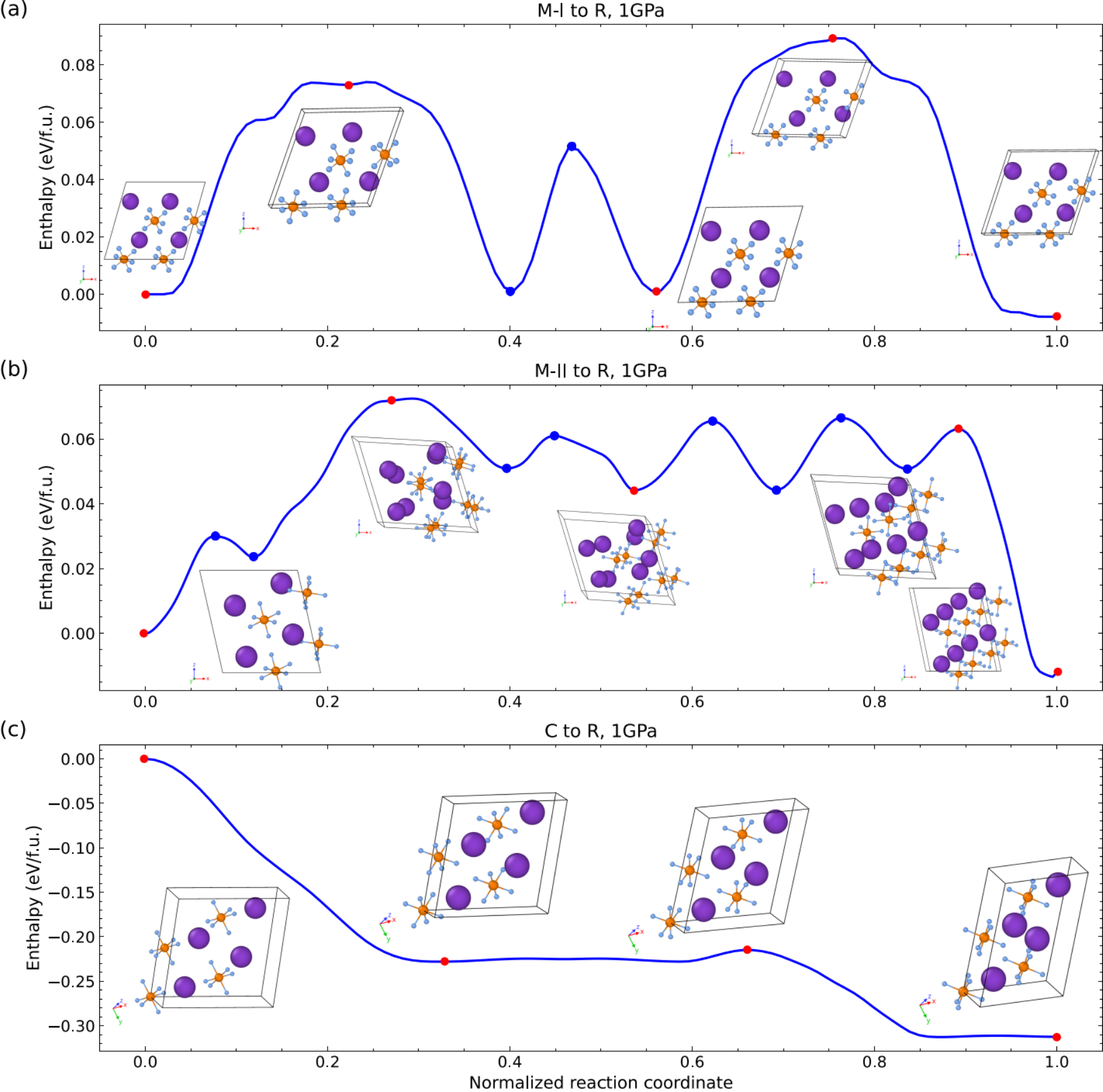}
\end{center}
\caption{Energy barriers for the pressure-induced phase transitions from the M-I, M-II, and C phases to the R phase, calculated at 0 K and 1 GPa using the MTP with the VCNEB method.
Insets display selected structures along the transformation pathways (corresponding to the images indicated by red points).
The complete phase transformation processes are available for visualization in the Supplementary Movies.}
\label{Fig7_VCNEB}
\end{figure*}

\begin{figure*}
\begin{center}
\includegraphics[width=0.7\textwidth,trim = {0.0cm 0.0cm 0.0cm 0.0cm}, clip]{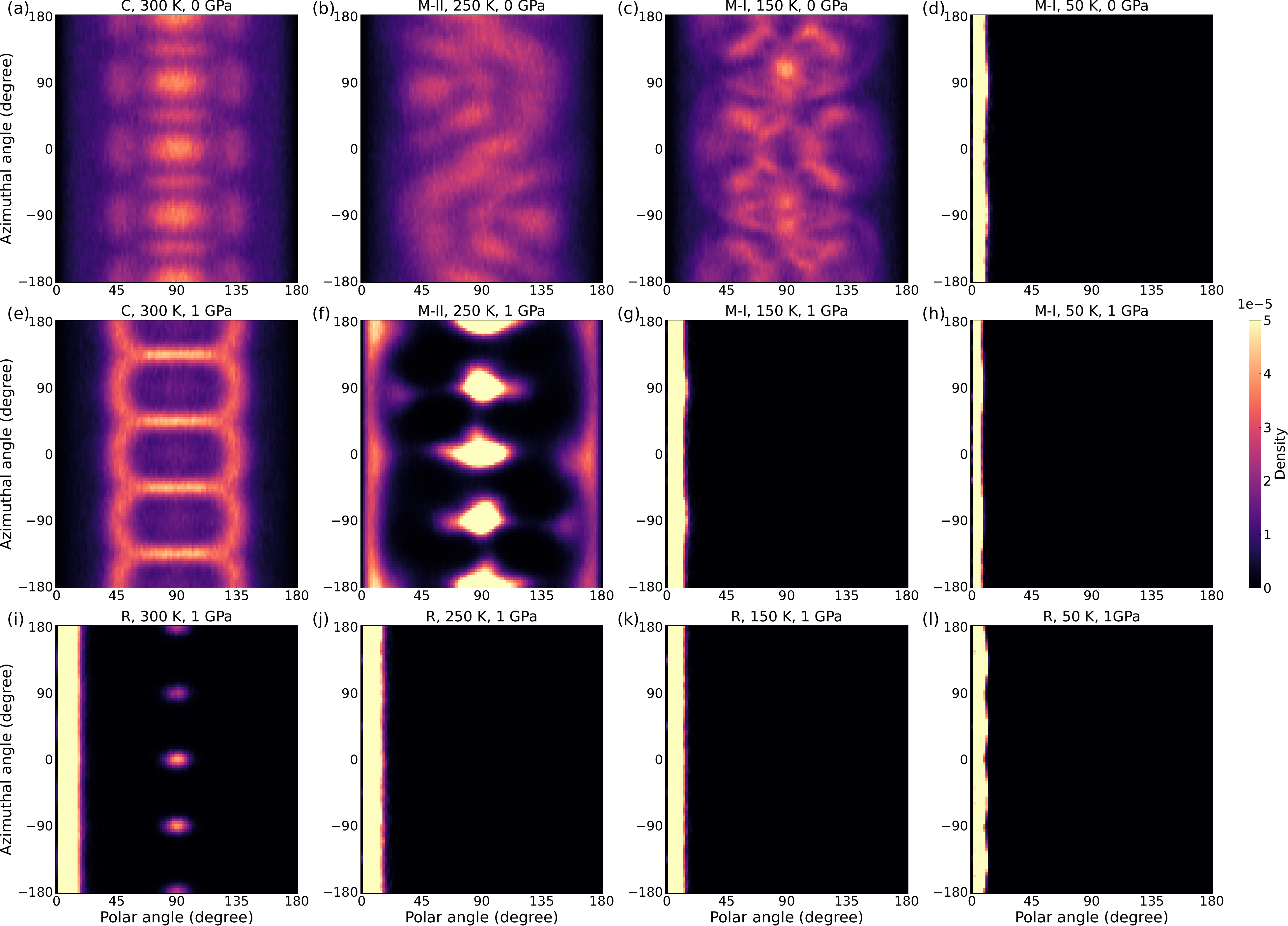}
\end{center}
\caption{Angular probability density for the PF bonds in the four phases of KPF$_6$ at specific temperature and pressure, obtained from the MTP-accelerated MD simulations.
The color coding highlights regions of high probability in light colors and low probability in dark colors.
}
\label{Fig8_orientational_disorder}
\end{figure*}

\begin{figure*}
\begin{center}
\includegraphics[width=0.95\textwidth,trim = {0.0cm 0.0cm 0.0cm 0.0cm}, clip]{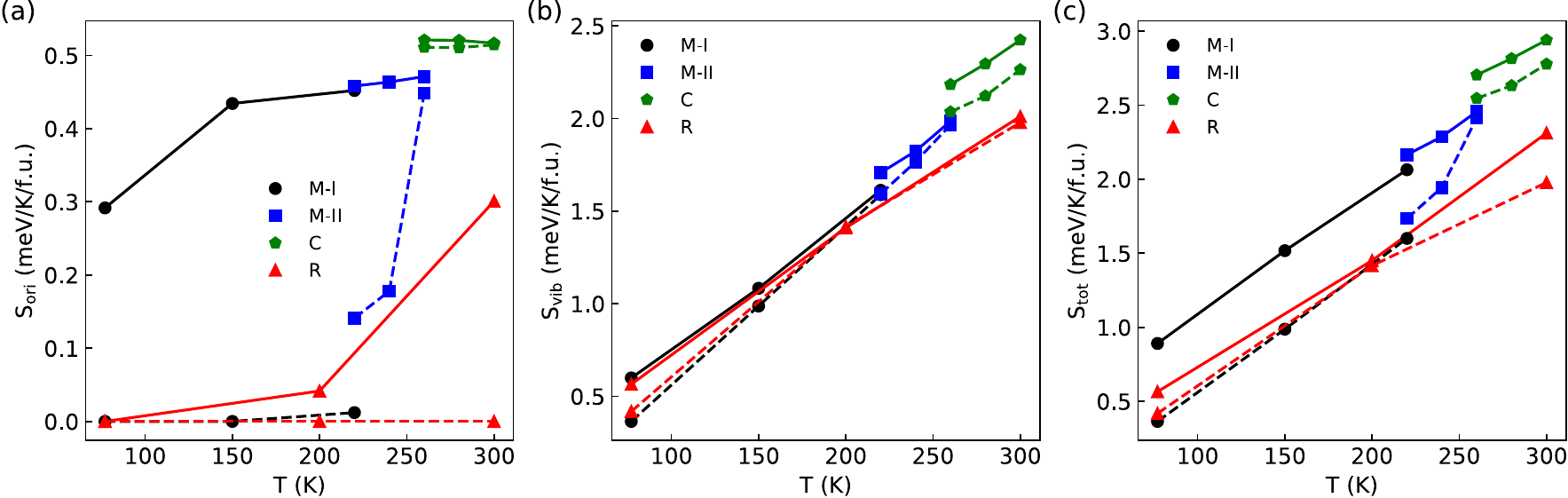}
\end{center}
\caption{MTP-predicted temperature-dependent entropies for the four phases of KPF$_6$ at 0 GPa (solid lines) and 1 GPa (dashed lines).
(a) Orientational entropy. (b) Vibrational entropy. (c) Total entropy.
}
\label{Fig9_entropy}
\end{figure*}

Experimentally, all M-I, M-II, and C phases undergo phase transformations to the R phase upon pressure~\cite{LibingNC2025}.
To estimate the energy barriers associated with these phase transitions,
we calculated the enthalpy at 0 K and 1 GPa along the respective transformation pathways using the VCNEB method~\cite{QIAN20132111}.
However, the VCNEB results are highly sensitive to the choice of the initial transformation pathway due to the facile rotation of the PF$_6$ octahedra, which can lead to slow convergence.
To address this problem, we utilized a strain-minimum transformation pathway identified through the Crystmatch code~\cite{wang2024crystal} as our initial pathway.
This approach resulted in minimal supercells containing 4, 8, and 4 formula units for the M-I, M-II, and C phases transitioning to the R phase, respectively.
To further enhance convergence efficiency, we partitioned the initial reaction pathway into multiple segments and performed independent VCNEB optimizations for each segment.
Convergence was deemed achieved when the maximum perpendicular forces fell below 0.02 eV/$\AA$.

The VCNEB results are displayed in Fig.~\ref{Fig7_VCNEB}.
One can observe the non-smooth transformation pathways manifested by the emergence of numerous metastable phases,
which is caused by the ease of PF$_6$  octahedral rotations coupled with lattice modulations.
To better visualize the detailed phase transformation process, we direct the reader to the Supplementary Movies.
Note that the exploration of such broad phase space during the energy barrier calculations has only been made possible by the developed accurate and efficient MTP.
The maximum energy barriers for the M-I and M-II phases transitioning to the R phase are found to be comparable, with values of 0.09 eV/f.u. and 0.07 eV/f.u., respectively.
These values are low for typical solid-solid phase transitions, indicating that these pressure-induced transitions can occur with relative ease.
Indeed, experimental studies have demonstrated that applying pressure in the range of just hundreds of MPa is sufficient to induce these phase transitions~\cite{LibingNC2025}.
It is crucial to highlight that the VCNEB method is a zero-temperature method that does not account for the vibrational and orientational disorder effects.
These effects are particularly significant for the high-temperature C phase as demonstrated in Fig.~\ref{Fig3_Lattice_C_NPT}.
Ignoring these effects would incorrectly result in a vanishing energy barrier for the transition from the C phase to the R phase [see Fig.~\ref{Fig7_VCNEB}(c)].

Despite the relatively low energy barriers in the pressure-induced phase transitions,
the dynamic phase transformation process has proven challenging to reproduce through direct unbiased MD simulations over time scales of hundreds of picoseconds.
This is largely due to the restricted time scales currently accessible in MD simulations, which may not adequately capture the dynamics of solid-solid phase transitions.
Therefore, we will focus our discussion on the  FOD and entropy of each phase at the specified temperature and pressure.
To this end, we computed the angular probability density for the PF bonds (see Fig.~\ref{Fig8_orientational_disorder})
and temperature-dependent entropies for the four phases of KPF$_6$ at 0 GPa and 1 GPa (Fig.~\ref{Fig9_entropy}).
Notably, the C phase at 300 K and 0 GPa exhibits the highest degree of FOD, characterized by predominant rotations
of the PF$_6$ octahedra [Fig.~\ref{Fig8_orientational_disorder}(a)].
In the M-II phase at 300 K and 0 GPa, we observe a noticeable yet slightly decreasing level of FOD [Fig.~\ref{Fig8_orientational_disorder}(b)].
Interestingly, even in the M-I phase at 150 K and 0 GPa, some degree of FOD persists [Fig.~\ref{Fig8_orientational_disorder}(c)],
which is only suppressed at the lower temperature of 50 K, where just local vibrations are present [Fig.~\ref{Fig8_orientational_disorder}(d)].
Applying a pressure of 1 GPa to the C, M-II, and M-I phases results in significant alterations to the fluorine orientation distributions,
leading to a reduction in FOD.
In particular, the FOD observed in the M-I phase at 150 K is completely suppressed under pressure [Fig.~\ref{Fig8_orientational_disorder}(g)].
In contrast, the high-pressure R phase exhibits a more ordered fluorine orientation, regardless of temperature variations [Figs.~\ref{Fig8_orientational_disorder}(i)-(l)],
while weak PF$_6$ octahedra rotations are only observed at 300 K [Fig.~\ref{Fig8_orientational_disorder}(i)].
These results underscore the significant role of FOD in the phase transitions of KPF$_6$.

The degree of FOD is quantitatively reflected by the predicted orientational entropy,
as illustrated in Fig.~\ref{Fig9_entropy}(a).
At 0 GPa, the orientational entropies decrease with the decreasing temperature, accompanied with the temperature-driven phase transitions from the C phase to the M-II phase to the M-I phase.
When a pressure of 1 GPa is applied, there is an overall reduction in orientational entropies.
This pressure effect is particularly pronounced in the M-I and R phases, where the FOD is nearly completely suppressed.
In comparison to the orientational entropy, vibrational entropy is considerably larger and thus dominates the total entropy, as shown in Fig.~\ref{Fig9_entropy}(b).
As expected, the vibrational entropy increases with temperature and decreases under pressure.
The total entropies including both orientational and vibrational contributions are displayed in Fig.~\ref{Fig9_entropy}(b).
Notably, within the considered temperature range of 77 to 300 K, there are persistent isothermal entropy changes upon pressure,
which are associated with pressure-induced phase transitions from the C, M-II, and M-I phases to the fully ordered R phase.
The isothermal entropy change is largest when transforming the C phase to the R phase under pressure,
attributed to the plastic crystal nature of the C phase, which results in nearly equal contributions from both orientational and vibrational entropy changes.
These findings align well with the experimental results, providing a comprehensive explanation for the observed all-temperature BCE in this material~\cite{LibingNC2025}.

\subsection{Electronic properties}

\begin{figure}
\begin{center}
\includegraphics[width=0.49\textwidth,trim = {0.0cm 0.0cm 0.0cm 0.0cm}, clip]{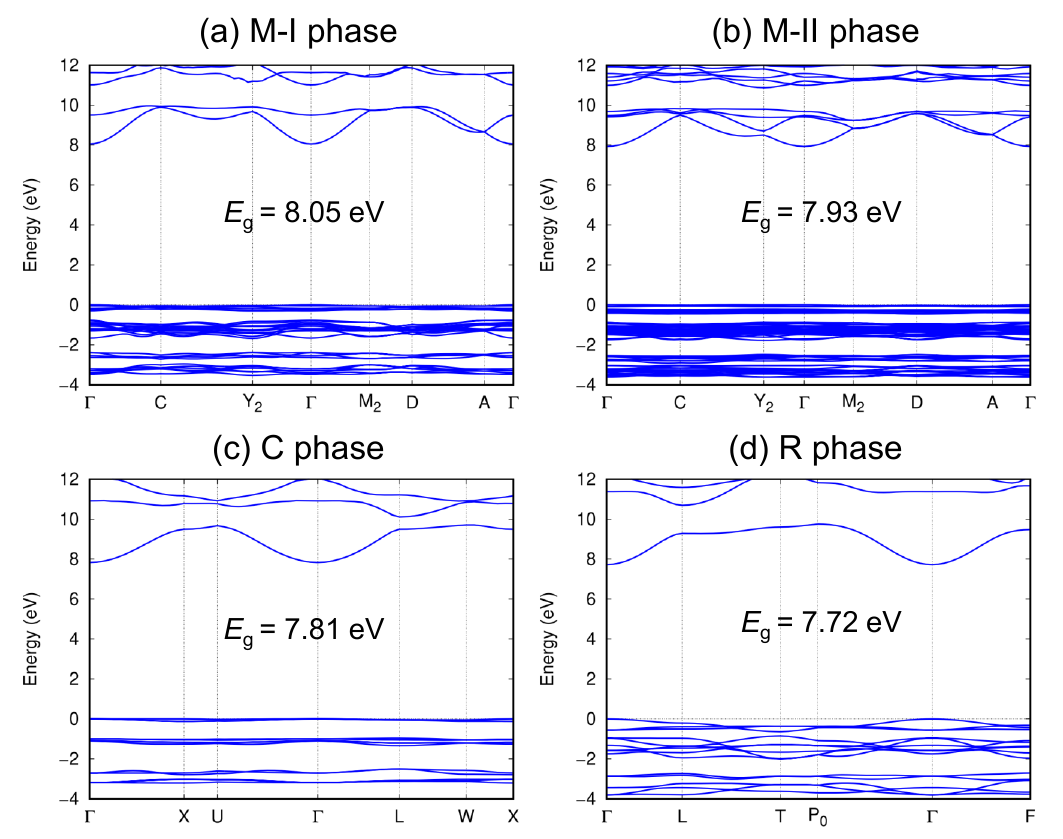}
\end{center}
\caption{Electronic band structures of the four phases of KPF$_6$ calculated using the HSE06 functional.
(a) The M-I phase. (b) The M-II phase. (c) The C phase. (d) The R phase.
The structures are obtained at 0 K and 0 GPa using the PBEsol functional.
}
\label{Fig10_band}
\end{figure}

Finally, we investigated the electronic properties of KPF$_6$.
Fig.~\ref{Fig10_band} presents the electronic band structures of the four phases of KPF$_6$ calculated using the HSE06 functional.
It is evident that all four phases exhibit a wide band gap of approximately 8 eV, indicating insulating behavior.
Such a large band gap results in low electrical conductivity, which is advantageous for practical applications of the BCE where efficient thermal management is crucial.
The valence bands are featured by the flat bands, while the conduction bands are more dispersive.
Atom-resolved electronic density of states analysis reveals that the bands near the gap are predominantly contributed by the fluorine atoms (Supplementary Information Fig.~S5~\cite{SM}),
highlighting their significant role also in the electronic properties of KPF$_6$.

\section{Conclusion}

In conclusion, this study has provided a comprehensive understanding of the atomistic mechanisms driving the phase transitions of KPF$_6$
through a combination of first-principles calculations and MLP-accelerated molecular dynamics simulations.
We have identified four distinct phases of KPF$_6$,
each characterized by its dynamic stability, phonon modes, and varying degrees of FOD.
Our findings indicate that the room-temperature C phase exhibits substantial FOD,
while the intermediate-temperature M-II and low-temperature M-I phases demonstrate a progressive reduction in FOD as the temperature decreases.
Notably, the application of pressure significantly suppresses the degree of FOD, leading to the emergence of the fully ordered R phase.
Our analysis highlights the critical role of thermal fluctuations in stabilizing the C phase, whereas the M-II, M-I, and R phases exhibit inherent stability at 0 K.
The interplay between FOD, anharmonic lattice dynamics, and cooperative octahedral rotations under pressure
enables KPF$_6$ to achieve persistent isothermal entropy changes across a broad temperature range through pressure-induced phase transitions.
These insights not only deepen our understanding of KPF$_6$,
but also pave the way for the design of advanced barocaloric materials with wide operating temperature spans.

\section*{Declaration of competing interest}
The authors declare that they have no known competing financial interests
or personal relationships that could have appeared to influence the work reported in this paper.

\section*{Author contributions}
P.L. conceived the project.
P.L. designed the research with the help of B.L. and X.-Q.C.
J.W. and P. L. performed the calculations.
P.L., B.L., and X.-Q.C. supervised the project.
Y.-C.Z., Y.L., H.D., M.L., and Y.S. participated in discussions.
P.L. wrote the manuscript with inputs from other authors.
All authors comments on the manuscript.

\section*{Acknowledgements}
This work is supported by
the National Natural Science Foundation of China (Grants No.~52422112, No.~52188101, and No.~52201030),
the Strategic Priority Research Program of the Chinese Academy of Sciences (XDA041040402),
the Liaoning Province Science and Technology Major Project (2024JH1/11700032, 2023021207-JH26/103 and RC230958),
the National Key R{\&}D Program of China 2021YFB3501503,
and
the Special Projects of the Central Government in Guidance of Local Science and Technology Development (2024010859-JH6/1006).

\bibliography{Reference}

\begin{thebibliography}{57}%
\makeatletter
\providecommand \@ifxundefined [1]{%
 \@ifx{#1\undefined}
}%
\providecommand \@ifnum [1]{%
 \ifnum #1\expandafter \@firstoftwo
 \else \expandafter \@secondoftwo
 \fi
}%
\providecommand \@ifx [1]{%
 \ifx #1\expandafter \@firstoftwo
 \else \expandafter \@secondoftwo
 \fi
}%
\providecommand \natexlab [1]{#1}%
\providecommand \enquote  [1]{``#1''}%
\providecommand \bibnamefont  [1]{#1}%
\providecommand \bibfnamefont [1]{#1}%
\providecommand \citenamefont [1]{#1}%
\providecommand \href@noop [0]{\@secondoftwo}%
\providecommand \href [0]{\begingroup \@sanitize@url \@href}%
\providecommand \@href[1]{\@@startlink{#1}\@@href}%
\providecommand \@@href[1]{\endgroup#1\@@endlink}%
\providecommand \@sanitize@url [0]{\catcode `\\12\catcode `\$12\catcode
  `\&12\catcode `\#12\catcode `\^12\catcode `\_12\catcode `\%12\relax}%
\providecommand \@@startlink[1]{}%
\providecommand \@@endlink[0]{}%
\providecommand \url  [0]{\begingroup\@sanitize@url \@url }%
\providecommand \@url [1]{\endgroup\@href {#1}{\urlprefix }}%
\providecommand \urlprefix  [0]{URL }%
\providecommand \Eprint [0]{\href }%
\providecommand \doibase [0]{http://dx.doi.org/}%
\providecommand \selectlanguage [0]{\@gobble}%
\providecommand \bibinfo  [0]{\@secondoftwo}%
\providecommand \bibfield  [0]{\@secondoftwo}%
\providecommand \translation [1]{[#1]}%
\providecommand \BibitemOpen [0]{}%
\providecommand \bibitemStop [0]{}%
\providecommand \bibitemNoStop [0]{.\EOS\space}%
\providecommand \EOS [0]{\spacefactor3000\relax}%
\providecommand \BibitemShut  [1]{\csname bibitem#1\endcsname}%
\let\auto@bib@innerbib\@empty
\bibitem [{\citenamefont {Shen}\ \emph {et~al.}(2009)\citenamefont {Shen},
  \citenamefont {Sun}, \citenamefont {Hu}, \citenamefont {Zhang},\ and\
  \citenamefont {Cheng}}]{Shen2009}%
  \BibitemOpen
  \bibfield  {author} {\bibinfo {author} {\bibfnamefont {B.~G.}\ \bibnamefont
  {Shen}}, \bibinfo {author} {\bibfnamefont {J.~R.}\ \bibnamefont {Sun}},
  \bibinfo {author} {\bibfnamefont {F.~X.}\ \bibnamefont {Hu}}, \bibinfo
  {author} {\bibfnamefont {H.~W.}\ \bibnamefont {Zhang}}, \ and\ \bibinfo
  {author} {\bibfnamefont {Z.~H.}\ \bibnamefont {Cheng}},\ }\bibfield  {title}
  {\enquote {\bibinfo {title} {Recent progress in exploring magnetocaloric
  materials},}\ }\href {\doibase 10.1002/adma.200901072} {\bibfield  {journal}
  {\bibinfo  {journal} {Advanced Materials}\ }\textbf {\bibinfo {volume}
  {21}},\ \bibinfo {pages} {4545--4564} (\bibinfo {year} {2009})}\BibitemShut
  {NoStop}%
\bibitem [{\citenamefont {Scott}(2011)}]{Scott2011}%
  \BibitemOpen
  \bibfield  {author} {\bibinfo {author} {\bibfnamefont {J.F.}\ \bibnamefont
  {Scott}},\ }\bibfield  {title} {\enquote {\bibinfo {title} {Electrocaloric
  materials},}\ }\href {\doibase
  https://doi.org/10.1146/annurev-matsci-062910-100341} {\bibfield  {journal}
  {\bibinfo  {journal} {Annual Review of Materials Research}\ }\textbf
  {\bibinfo {volume} {41}},\ \bibinfo {pages} {229--240} (\bibinfo {year}
  {2011})}\BibitemShut {NoStop}%
\bibitem [{\citenamefont {Ma{\~{n}}osa}\ and\ \citenamefont
  {Planes}(2017)}]{Mechanocaloric2017}%
  \BibitemOpen
  \bibfield  {author} {\bibinfo {author} {\bibfnamefont {Llu{\'i}s}\
  \bibnamefont {Ma{\~{n}}osa}}\ and\ \bibinfo {author} {\bibfnamefont {Antoni}\
  \bibnamefont {Planes}},\ }\bibfield  {title} {\enquote {\bibinfo {title}
  {Materials with giant mechanocaloric effects: Cooling by strength},}\ }\href
  {\doibase 10.1002/adma.201603607} {\bibfield  {journal} {\bibinfo  {journal}
  {Advanced Materials}\ }\textbf {\bibinfo {volume} {29}},\ \bibinfo {pages}
  {1603607} (\bibinfo {year} {2017})}\BibitemShut {NoStop}%
\bibitem [{\citenamefont {Ma{\~{n}}osa}\ \emph {et~al.}(2010)\citenamefont
  {Ma{\~{n}}osa}, \citenamefont {Gonz{\'a}lez-Alonso}, \citenamefont {Planes},
  \citenamefont {Bonnot}, \citenamefont {Barrio}, \citenamefont {Tamarit},
  \citenamefont {Aksoy},\ and\ \citenamefont {Acet}}]{barocaloric2010}%
  \BibitemOpen
  \bibfield  {author} {\bibinfo {author} {\bibfnamefont {Llu{\'i}s}\
  \bibnamefont {Ma{\~{n}}osa}}, \bibinfo {author} {\bibfnamefont {David}\
  \bibnamefont {Gonz{\'a}lez-Alonso}}, \bibinfo {author} {\bibfnamefont
  {Antoni}\ \bibnamefont {Planes}}, \bibinfo {author} {\bibfnamefont {Erell}\
  \bibnamefont {Bonnot}}, \bibinfo {author} {\bibfnamefont {Maria}\
  \bibnamefont {Barrio}}, \bibinfo {author} {\bibfnamefont {Josep-Llu{\'i}s}\
  \bibnamefont {Tamarit}}, \bibinfo {author} {\bibfnamefont {Seda}\
  \bibnamefont {Aksoy}}, \ and\ \bibinfo {author} {\bibfnamefont {Mehmet}\
  \bibnamefont {Acet}},\ }\bibfield  {title} {\enquote {\bibinfo {title} {Giant
  solid-state barocaloric effect in the {{Ni}}--{{Mn}}--{{In}} magnetic
  shape-memory alloy},}\ }\href {\doibase 10.1038/nmat2731} {\bibfield
  {journal} {\bibinfo  {journal} {Nature Materials}\ }\textbf {\bibinfo
  {volume} {9}},\ \bibinfo {pages} {478--481} (\bibinfo {year}
  {2010})}\BibitemShut {NoStop}%
\bibitem [{\citenamefont {Cirillo}\ \emph {et~al.}(2022)\citenamefont
  {Cirillo}, \citenamefont {Greco},\ and\ \citenamefont
  {Masselli}}]{Review_Cirillo2022}%
  \BibitemOpen
  \bibfield  {author} {\bibinfo {author} {\bibfnamefont {Luca}\ \bibnamefont
  {Cirillo}}, \bibinfo {author} {\bibfnamefont {Adriana}\ \bibnamefont
  {Greco}}, \ and\ \bibinfo {author} {\bibfnamefont {Claudia}\ \bibnamefont
  {Masselli}},\ }\bibfield  {title} {\enquote {\bibinfo {title} {Cooling
  through barocaloric effect: A review of the state of the art up to 2022},}\
  }\href {https://www.sciencedirect.com/science/article/pii/S245190492200186X}
  {\bibfield  {journal} {\bibinfo  {journal} {Thermal Science and Engineering
  Progress}\ }\textbf {\bibinfo {volume} {33}},\ \bibinfo {pages} {101380}
  (\bibinfo {year} {2022})}\BibitemShut {NoStop}%
\bibitem [{\citenamefont {Aznar}\ \emph {et~al.}(2019)\citenamefont {Aznar},
  \citenamefont {Gr{\`a}cia-Condal}, \citenamefont {Planes}, \citenamefont
  {Lloveras}, \citenamefont {Barrio}, \citenamefont {Tamarit}, \citenamefont
  {Xiong}, \citenamefont {Cong}, \citenamefont {Popescu},\ and\ \citenamefont
  {Ma{\~{n}}osa}}]{PhysRevMaterials.3.044406}%
  \BibitemOpen
  \bibfield  {author} {\bibinfo {author} {\bibfnamefont {Araceli}\ \bibnamefont
  {Aznar}}, \bibinfo {author} {\bibfnamefont {Adri{\`a}}\ \bibnamefont
  {Gr{\`a}cia-Condal}}, \bibinfo {author} {\bibfnamefont {Antoni}\ \bibnamefont
  {Planes}}, \bibinfo {author} {\bibfnamefont {Pol}\ \bibnamefont {Lloveras}},
  \bibinfo {author} {\bibfnamefont {Maria}\ \bibnamefont {Barrio}}, \bibinfo
  {author} {\bibfnamefont {Josep-Llu{\'i}s}\ \bibnamefont {Tamarit}}, \bibinfo
  {author} {\bibfnamefont {Wenxin}\ \bibnamefont {Xiong}}, \bibinfo {author}
  {\bibfnamefont {Daoyong}\ \bibnamefont {Cong}}, \bibinfo {author}
  {\bibfnamefont {Catalin}\ \bibnamefont {Popescu}}, \ and\ \bibinfo {author}
  {\bibfnamefont {Llu{\'i}s}\ \bibnamefont {Ma{\~{n}}osa}},\ }\bibfield
  {title} {\enquote {\bibinfo {title} {Giant barocaloric effect in
  all-$d$-metal heusler shape memory alloys},}\ }\href {\doibase
  10.1103/PhysRevMaterials.3.044406} {\bibfield  {journal} {\bibinfo  {journal}
  {Physical Review Materials}\ }\textbf {\bibinfo {volume} {3}},\ \bibinfo
  {pages} {044406} (\bibinfo {year} {2019})}\BibitemShut {NoStop}%
\bibitem [{\citenamefont {Stern-Taulats}\ \emph {et~al.}(2015)\citenamefont
  {Stern-Taulats}, \citenamefont {Planes}, \citenamefont {Lloveras},
  \citenamefont {Barrio}, \citenamefont {Tamarit}, \citenamefont {Pramanick},
  \citenamefont {Majumdar}, \citenamefont {Y{\"u}ce}, \citenamefont {Emre},
  \citenamefont {Frontera},\ and\ \citenamefont
  {Ma{\~{n}}osa}}]{Stern-Taulats2015}%
  \BibitemOpen
  \bibfield  {author} {\bibinfo {author} {\bibfnamefont {Enric}\ \bibnamefont
  {Stern-Taulats}}, \bibinfo {author} {\bibfnamefont {Antoni}\ \bibnamefont
  {Planes}}, \bibinfo {author} {\bibfnamefont {Pol}\ \bibnamefont {Lloveras}},
  \bibinfo {author} {\bibfnamefont {Maria}\ \bibnamefont {Barrio}}, \bibinfo
  {author} {\bibfnamefont {Josep-Llu{\'i}s}\ \bibnamefont {Tamarit}}, \bibinfo
  {author} {\bibfnamefont {Sabyasachi}\ \bibnamefont {Pramanick}}, \bibinfo
  {author} {\bibfnamefont {Subham}\ \bibnamefont {Majumdar}}, \bibinfo {author}
  {\bibfnamefont {Suheyla}\ \bibnamefont {Y{\"u}ce}}, \bibinfo {author}
  {\bibfnamefont {Baris}\ \bibnamefont {Emre}}, \bibinfo {author}
  {\bibfnamefont {Carlos}\ \bibnamefont {Frontera}}, \ and\ \bibinfo {author}
  {\bibfnamefont {Llu{\'i}s}\ \bibnamefont {Ma{\~{n}}osa}},\ }\bibfield
  {title} {\enquote {\bibinfo {title} {Tailoring barocaloric and magnetocaloric
  properties in low-hysteresis magnetic shape memory alloys},}\ }\href
  {https://www.sciencedirect.com/science/article/pii/S1359645415004127}
  {\bibfield  {journal} {\bibinfo  {journal} {Acta Materialia}\ }\textbf
  {\bibinfo {volume} {96}},\ \bibinfo {pages} {324--332} (\bibinfo {year}
  {2015})}\BibitemShut {NoStop}%
\bibitem [{\citenamefont {Cazorla}\ and\ \citenamefont
  {Errandonea}(2016)}]{Cazorla2016}%
  \BibitemOpen
  \bibfield  {author} {\bibinfo {author} {\bibfnamefont {Claudio}\ \bibnamefont
  {Cazorla}}\ and\ \bibinfo {author} {\bibfnamefont {Daniel}\ \bibnamefont
  {Errandonea}},\ }\bibfield  {title} {\enquote {\bibinfo {title} {Giant
  mechanocaloric effects in fluorite-structured superionic materials},}\ }\href
  {\doibase 10.1021/acs.nanolett.6b00422} {\bibfield  {journal} {\bibinfo
  {journal} {Nano Letters}\ }\textbf {\bibinfo {volume} {16}},\ \bibinfo
  {pages} {3124--3129} (\bibinfo {year} {2016})}\BibitemShut {NoStop}%
\bibitem [{\citenamefont {Aznar}\ \emph {et~al.}(2017)\citenamefont {Aznar},
  \citenamefont {Lloveras}, \citenamefont {Romanini}, \citenamefont {Barrio},
  \citenamefont {Tamarit}, \citenamefont {Cazorla}, \citenamefont {Errandonea},
  \citenamefont {Mathur}, \citenamefont {Planes}, \citenamefont {Moya},\ and\
  \citenamefont {Ma{\~{n}}osa}}]{AgI_NC2017}%
  \BibitemOpen
  \bibfield  {author} {\bibinfo {author} {\bibfnamefont {Araceli}\ \bibnamefont
  {Aznar}}, \bibinfo {author} {\bibfnamefont {Pol}\ \bibnamefont {Lloveras}},
  \bibinfo {author} {\bibfnamefont {Michela}\ \bibnamefont {Romanini}},
  \bibinfo {author} {\bibfnamefont {Mar{\'i}a}\ \bibnamefont {Barrio}},
  \bibinfo {author} {\bibfnamefont {Josep-Llu{\'i}s}\ \bibnamefont {Tamarit}},
  \bibinfo {author} {\bibfnamefont {Claudio}\ \bibnamefont {Cazorla}}, \bibinfo
  {author} {\bibfnamefont {Daniel}\ \bibnamefont {Errandonea}}, \bibinfo
  {author} {\bibfnamefont {Neil~D.}\ \bibnamefont {Mathur}}, \bibinfo {author}
  {\bibfnamefont {Antoni}\ \bibnamefont {Planes}}, \bibinfo {author}
  {\bibfnamefont {Xavier}\ \bibnamefont {Moya}}, \ and\ \bibinfo {author}
  {\bibfnamefont {Llu{\'i}s}\ \bibnamefont {Ma{\~{n}}osa}},\ }\bibfield
  {title} {\enquote {\bibinfo {title} {Giant barocaloric effects over a wide
  temperature range in superionic conductor {{AgI}}},}\ }\href {\doibase
  10.1038/s41467-017-01898-2} {\bibfield  {journal} {\bibinfo  {journal}
  {Nature Communications}\ }\textbf {\bibinfo {volume} {8}},\ \bibinfo {pages}
  {1851} (\bibinfo {year} {2017})}\BibitemShut {NoStop}%
\bibitem [{\citenamefont {Min}\ \emph {et~al.}(2020)\citenamefont {Min},
  \citenamefont {Sagotra},\ and\ \citenamefont
  {Cazorla}}]{PhysRevMaterials.4.015403}%
  \BibitemOpen
  \bibfield  {author} {\bibinfo {author} {\bibfnamefont {Jie}\ \bibnamefont
  {Min}}, \bibinfo {author} {\bibfnamefont {Arun~K.}\ \bibnamefont {Sagotra}},
  \ and\ \bibinfo {author} {\bibfnamefont {Claudio}\ \bibnamefont {Cazorla}},\
  }\bibfield  {title} {\enquote {\bibinfo {title} {Large barocaloric effects in
  thermoelectric superionic materials},}\ }\href {\doibase
  10.1103/PhysRevMaterials.4.015403} {\bibfield  {journal} {\bibinfo  {journal}
  {Phys. Rev. Mater.}\ }\textbf {\bibinfo {volume} {4}},\ \bibinfo {pages}
  {015403} (\bibinfo {year} {2020})}\BibitemShut {NoStop}%
\bibitem [{\citenamefont {Berm{\'u}dez-Garc{\'i}a}\ \emph
  {et~al.}(2017{\natexlab{a}})\citenamefont {Berm{\'u}dez-Garc{\'i}a},
  \citenamefont {S{\'a}nchez-And{\'u}jar}, \citenamefont {Castro-Garc{\'i}a},
  \citenamefont {L{\'o}pez-Beceiro}, \citenamefont {Artiaga},\ and\
  \citenamefont {Se{\~{n}}ar{\'i}s-Rodr{\'i}guez}}]{hybridperovskite_NC2017}%
  \BibitemOpen
  \bibfield  {author} {\bibinfo {author} {\bibfnamefont {Juan~M.}\ \bibnamefont
  {Berm{\'u}dez-Garc{\'i}a}}, \bibinfo {author} {\bibfnamefont {Manuel}\
  \bibnamefont {S{\'a}nchez-And{\'u}jar}}, \bibinfo {author} {\bibfnamefont
  {Socorro}\ \bibnamefont {Castro-Garc{\'i}a}}, \bibinfo {author}
  {\bibfnamefont {Jorge}\ \bibnamefont {L{\'o}pez-Beceiro}}, \bibinfo {author}
  {\bibfnamefont {Ram{\'o}n}\ \bibnamefont {Artiaga}}, \ and\ \bibinfo {author}
  {\bibfnamefont {Mar{\'i}a~A.}\ \bibnamefont
  {Se{\~{n}}ar{\'i}s-Rodr{\'i}guez}},\ }\bibfield  {title} {\enquote {\bibinfo
  {title} {Giant barocaloric effect in the ferroic organic-inorganic hybrid
  {{[TPrA][Mn(dca)3]}} perovskite under easily accessible pressures},}\ }\href
  {\doibase 10.1038/ncomms15715} {\bibfield  {journal} {\bibinfo  {journal}
  {Nature Communications}\ }\textbf {\bibinfo {volume} {8}},\ \bibinfo {pages}
  {15715} (\bibinfo {year} {2017}{\natexlab{a}})}\BibitemShut {NoStop}%
\bibitem [{\citenamefont {Berm{\'u}dez-Garc{\'i}a}\ \emph
  {et~al.}(2017{\natexlab{b}})\citenamefont {Berm{\'u}dez-Garc{\'i}a},
  \citenamefont {S{\'a}nchez-And{\'u}jar},\ and\ \citenamefont
  {Se{\~{n}}ar{\'i}s-Rodr{\'i}guez}}]{JPCL2017}%
  \BibitemOpen
  \bibfield  {author} {\bibinfo {author} {\bibfnamefont {Juan~M.}\ \bibnamefont
  {Berm{\'u}dez-Garc{\'i}a}}, \bibinfo {author} {\bibfnamefont {Manuel}\
  \bibnamefont {S{\'a}nchez-And{\'u}jar}}, \ and\ \bibinfo {author}
  {\bibfnamefont {Mar{\'i}a~A.}\ \bibnamefont
  {Se{\~{n}}ar{\'i}s-Rodr{\'i}guez}},\ }\bibfield  {title} {\enquote {\bibinfo
  {title} {A new playground for organic--inorganic hybrids: Barocaloric
  materials for pressure-induced solid-state cooling},}\ }\href {\doibase
  10.1021/acs.jpclett.7b01845} {\bibfield  {journal} {\bibinfo  {journal} {The
  Journal of Physical Chemistry Letters}\ }\textbf {\bibinfo {volume} {8}},\
  \bibinfo {pages} {4419--4423} (\bibinfo {year}
  {2017}{\natexlab{b}})}\BibitemShut {NoStop}%
\bibitem [{\citenamefont {Salgado-Beceiro}\ \emph {et~al.}(2020)\citenamefont
  {Salgado-Beceiro}, \citenamefont {Nonato}, \citenamefont {Silva},
  \citenamefont {Garc{\'i}a-Fern{\'a}ndez}, \citenamefont
  {S{\'a}nchez-And{\'u}jar}, \citenamefont {Castro-Garc{\'i}a}, \citenamefont
  {Stern-Taulats}, \citenamefont {Se{\~{n}}ar{\'i}s-Rodr{\'i}guez},
  \citenamefont {Moya},\ and\ \citenamefont
  {Berm{\'u}dez-Garc{\'i}a}}]{Salgado-Beceiro2020}%
  \BibitemOpen
  \bibfield  {author} {\bibinfo {author} {\bibfnamefont {Jorge}\ \bibnamefont
  {Salgado-Beceiro}}, \bibinfo {author} {\bibfnamefont {Ariel}\ \bibnamefont
  {Nonato}}, \bibinfo {author} {\bibfnamefont {Rosivaldo~Xavier}\ \bibnamefont
  {Silva}}, \bibinfo {author} {\bibfnamefont {Alberto}\ \bibnamefont
  {Garc{\'i}a-Fern{\'a}ndez}}, \bibinfo {author} {\bibfnamefont {Manuel}\
  \bibnamefont {S{\'a}nchez-And{\'u}jar}}, \bibinfo {author} {\bibfnamefont
  {Socorro}\ \bibnamefont {Castro-Garc{\'i}a}}, \bibinfo {author}
  {\bibfnamefont {Enric}\ \bibnamefont {Stern-Taulats}}, \bibinfo {author}
  {\bibfnamefont {Mar{\'i}a~Antonia}\ \bibnamefont
  {Se{\~{n}}ar{\'i}s-Rodr{\'i}guez}}, \bibinfo {author} {\bibfnamefont
  {Xavier}\ \bibnamefont {Moya}}, \ and\ \bibinfo {author} {\bibfnamefont
  {Juan~Manuel}\ \bibnamefont {Berm{\'u}dez-Garc{\'i}a}},\ }\bibfield  {title}
  {\enquote {\bibinfo {title} {Near-room-temperature reversible giant
  barocaloric effects in {{[(CH3)4N]Mn[N3]3}} hybrid perovskite},}\ }\href
  {https://www.sciencedirect.com/science/article/pii/S263354092300508X}
  {\bibfield  {journal} {\bibinfo  {journal} {Materials Advances}\ }\textbf
  {\bibinfo {volume} {1}},\ \bibinfo {pages} {3167--3170} (\bibinfo {year}
  {2020})}\BibitemShut {NoStop}%
\bibitem [{\citenamefont {Li}\ \emph {et~al.}(2021)\citenamefont {Li},
  \citenamefont {Barrio}, \citenamefont {Dunstan}, \citenamefont {Dixey},
  \citenamefont {Lou}, \citenamefont {Tamarit}, \citenamefont {Phillips},\ and\
  \citenamefont {Lloveras}}]{LiAFM2021}%
  \BibitemOpen
  \bibfield  {author} {\bibinfo {author} {\bibfnamefont {Junning}\ \bibnamefont
  {Li}}, \bibinfo {author} {\bibfnamefont {Mar{\'i}a}\ \bibnamefont {Barrio}},
  \bibinfo {author} {\bibfnamefont {David~J.}\ \bibnamefont {Dunstan}},
  \bibinfo {author} {\bibfnamefont {Richard}\ \bibnamefont {Dixey}}, \bibinfo
  {author} {\bibfnamefont {Xiaojie}\ \bibnamefont {Lou}}, \bibinfo {author}
  {\bibfnamefont {Josep-Llu{\'i}s}\ \bibnamefont {Tamarit}}, \bibinfo {author}
  {\bibfnamefont {Anthony~E.}\ \bibnamefont {Phillips}}, \ and\ \bibinfo
  {author} {\bibfnamefont {Pol}\ \bibnamefont {Lloveras}},\ }\bibfield  {title}
  {\enquote {\bibinfo {title} {Colossal reversible barocaloric effects in
  layered hybrid perovskite {{(C10H21NH3)2MnCl4}} under low pressure near room
  temperature},}\ }\href {\doibase 10.1002/adfm.202105154} {\bibfield
  {journal} {\bibinfo  {journal} {Advanced Functional Materials}\ }\textbf
  {\bibinfo {volume} {31}},\ \bibinfo {pages} {2105154} (\bibinfo {year}
  {2021})}\BibitemShut {NoStop}%
\bibitem [{\citenamefont {Li}\ \emph {et~al.}(2019)\citenamefont {Li},
  \citenamefont {Kawakita}, \citenamefont {Ohira-Kawamura}, \citenamefont
  {Sugahara}, \citenamefont {Wang}, \citenamefont {Wang}, \citenamefont {Chen},
  \citenamefont {Kawaguchi}, \citenamefont {Kawaguchi}, \citenamefont {Ohara},
  \citenamefont {Li}, \citenamefont {Yu}, \citenamefont {Mole}, \citenamefont
  {Hattori}, \citenamefont {Kikuchi}, \citenamefont {Yano}, \citenamefont
  {Zhang}, \citenamefont {Zhang}, \citenamefont {Ren}, \citenamefont {Lin},
  \citenamefont {Sakata}, \citenamefont {Nakajima},\ and\ \citenamefont
  {Zhang}}]{Li_Nature2019}%
  \BibitemOpen
  \bibfield  {author} {\bibinfo {author} {\bibfnamefont {Bing}\ \bibnamefont
  {Li}}, \bibinfo {author} {\bibfnamefont {Yukinobu}\ \bibnamefont {Kawakita}},
  \bibinfo {author} {\bibfnamefont {Seiko}\ \bibnamefont {Ohira-Kawamura}},
  \bibinfo {author} {\bibfnamefont {Takeshi}\ \bibnamefont {Sugahara}},
  \bibinfo {author} {\bibfnamefont {Hui}\ \bibnamefont {Wang}}, \bibinfo
  {author} {\bibfnamefont {Jingfan}\ \bibnamefont {Wang}}, \bibinfo {author}
  {\bibfnamefont {Yanna}\ \bibnamefont {Chen}}, \bibinfo {author}
  {\bibfnamefont {Saori~I.}\ \bibnamefont {Kawaguchi}}, \bibinfo {author}
  {\bibfnamefont {Shogo}\ \bibnamefont {Kawaguchi}}, \bibinfo {author}
  {\bibfnamefont {Koji}\ \bibnamefont {Ohara}}, \bibinfo {author}
  {\bibfnamefont {Kuo}\ \bibnamefont {Li}}, \bibinfo {author} {\bibfnamefont
  {Dehong}\ \bibnamefont {Yu}}, \bibinfo {author} {\bibfnamefont {Richard}\
  \bibnamefont {Mole}}, \bibinfo {author} {\bibfnamefont {Takanori}\
  \bibnamefont {Hattori}}, \bibinfo {author} {\bibfnamefont {Tatsuya}\
  \bibnamefont {Kikuchi}}, \bibinfo {author} {\bibfnamefont {Shin-ichiro}\
  \bibnamefont {Yano}}, \bibinfo {author} {\bibfnamefont {Zhao}\ \bibnamefont
  {Zhang}}, \bibinfo {author} {\bibfnamefont {Zhe}\ \bibnamefont {Zhang}},
  \bibinfo {author} {\bibfnamefont {Weijun}\ \bibnamefont {Ren}}, \bibinfo
  {author} {\bibfnamefont {Shangchao}\ \bibnamefont {Lin}}, \bibinfo {author}
  {\bibfnamefont {Osami}\ \bibnamefont {Sakata}}, \bibinfo {author}
  {\bibfnamefont {Kenji}\ \bibnamefont {Nakajima}}, \ and\ \bibinfo {author}
  {\bibfnamefont {Zhidong}\ \bibnamefont {Zhang}},\ }\bibfield  {title}
  {\enquote {\bibinfo {title} {Colossal barocaloric effects in plastic
  crystals},}\ }\href {\doibase 10.1038/s41586-019-1042-5} {\bibfield
  {journal} {\bibinfo  {journal} {Nature}\ }\textbf {\bibinfo {volume} {567}},\
  \bibinfo {pages} {506--510} (\bibinfo {year} {2019})}\BibitemShut {NoStop}%
\bibitem [{\citenamefont {Lloveras}\ \emph {et~al.}(2019)\citenamefont
  {Lloveras}, \citenamefont {Aznar}, \citenamefont {Barrio}, \citenamefont
  {Negrier}, \citenamefont {Popescu}, \citenamefont {Planes}, \citenamefont
  {Ma{\~{n}}osa}, \citenamefont {Stern-Taulats}, \citenamefont {Avramenko},
  \citenamefont {Mathur}, \citenamefont {Moya},\ and\ \citenamefont
  {Tamarit}}]{Lloveras2019}%
  \BibitemOpen
  \bibfield  {author} {\bibinfo {author} {\bibfnamefont {P.}~\bibnamefont
  {Lloveras}}, \bibinfo {author} {\bibfnamefont {A.}~\bibnamefont {Aznar}},
  \bibinfo {author} {\bibfnamefont {M.}~\bibnamefont {Barrio}}, \bibinfo
  {author} {\bibfnamefont {Ph.}\ \bibnamefont {Negrier}}, \bibinfo {author}
  {\bibfnamefont {C.}~\bibnamefont {Popescu}}, \bibinfo {author} {\bibfnamefont
  {A.}~\bibnamefont {Planes}}, \bibinfo {author} {\bibfnamefont
  {L.}~\bibnamefont {Ma{\~{n}}osa}}, \bibinfo {author} {\bibfnamefont
  {E.}~\bibnamefont {Stern-Taulats}}, \bibinfo {author} {\bibfnamefont
  {A.}~\bibnamefont {Avramenko}}, \bibinfo {author} {\bibfnamefont {N.~D.}\
  \bibnamefont {Mathur}}, \bibinfo {author} {\bibfnamefont {X.}~\bibnamefont
  {Moya}}, \ and\ \bibinfo {author} {\bibfnamefont {J.-Ll.}\ \bibnamefont
  {Tamarit}},\ }\bibfield  {title} {\enquote {\bibinfo {title} {Colossal
  barocaloric effects near room temperature in plastic crystals of
  neopentylglycol},}\ }\href {\doibase 10.1038/s41467-019-09730-9} {\bibfield
  {journal} {\bibinfo  {journal} {Nature Communications}\ }\textbf {\bibinfo
  {volume} {10}},\ \bibinfo {pages} {1803} (\bibinfo {year}
  {2019})}\BibitemShut {NoStop}%
\bibitem [{\citenamefont {Aznar}\ \emph {et~al.}(2020)\citenamefont {Aznar},
  \citenamefont {Lloveras}, \citenamefont {Barrio}, \citenamefont {Negrier},
  \citenamefont {Planes}, \citenamefont {Ma{\~{n}}osa}, \citenamefont {Mathur},
  \citenamefont {Moya},\ and\ \citenamefont {Tamarit}}]{Aznar2020}%
  \BibitemOpen
  \bibfield  {author} {\bibinfo {author} {\bibfnamefont {Araceli}\ \bibnamefont
  {Aznar}}, \bibinfo {author} {\bibfnamefont {Pol}\ \bibnamefont {Lloveras}},
  \bibinfo {author} {\bibfnamefont {Mar{\'i}a}\ \bibnamefont {Barrio}},
  \bibinfo {author} {\bibfnamefont {Philippe}\ \bibnamefont {Negrier}},
  \bibinfo {author} {\bibfnamefont {Antoni}\ \bibnamefont {Planes}}, \bibinfo
  {author} {\bibfnamefont {Llu{\'i}s}\ \bibnamefont {Ma{\~{n}}osa}}, \bibinfo
  {author} {\bibfnamefont {Neil~D.}\ \bibnamefont {Mathur}}, \bibinfo {author}
  {\bibfnamefont {Xavier}\ \bibnamefont {Moya}}, \ and\ \bibinfo {author}
  {\bibfnamefont {Josep-Llu{\'i}s}\ \bibnamefont {Tamarit}},\ }\bibfield
  {title} {\enquote {\bibinfo {title} {Reversible and irreversible colossal
  barocaloric effects in plastic crystals},}\ }\href {\doibase
  10.1039/C9TA10947A} {\bibfield  {journal} {\bibinfo  {journal} {Journal of
  Materials Chemistry A}\ }\textbf {\bibinfo {volume} {8}},\ \bibinfo {pages}
  {639--647} (\bibinfo {year} {2020})}\BibitemShut {NoStop}%
\bibitem [{\citenamefont {Li}\ \emph {et~al.}(2020)\citenamefont {Li},
  \citenamefont {Li}, \citenamefont {Xu}, \citenamefont {Yang}, \citenamefont
  {Xu}, \citenamefont {Jia}, \citenamefont {Li}, \citenamefont {He},
  \citenamefont {Li},\ and\ \citenamefont {Wang}}]{Wanghui_NC2020}%
  \BibitemOpen
  \bibfield  {author} {\bibinfo {author} {\bibfnamefont {F.~B.}\ \bibnamefont
  {Li}}, \bibinfo {author} {\bibfnamefont {M.}~\bibnamefont {Li}}, \bibinfo
  {author} {\bibfnamefont {X.}~\bibnamefont {Xu}}, \bibinfo {author}
  {\bibfnamefont {Z.~C.}\ \bibnamefont {Yang}}, \bibinfo {author}
  {\bibfnamefont {H.}~\bibnamefont {Xu}}, \bibinfo {author} {\bibfnamefont
  {C.~K.}\ \bibnamefont {Jia}}, \bibinfo {author} {\bibfnamefont
  {K.}~\bibnamefont {Li}}, \bibinfo {author} {\bibfnamefont {J.}~\bibnamefont
  {He}}, \bibinfo {author} {\bibfnamefont {B.}~\bibnamefont {Li}}, \ and\
  \bibinfo {author} {\bibfnamefont {Hui}\ \bibnamefont {Wang}},\ }\bibfield
  {title} {\enquote {\bibinfo {title} {Understanding colossal barocaloric
  effects in plastic crystals},}\ }\href {\doibase 10.1038/s41467-020-18043-1}
  {\bibfield  {journal} {\bibinfo  {journal} {Nature Communications}\ }\textbf
  {\bibinfo {volume} {11}},\ \bibinfo {pages} {4190} (\bibinfo {year}
  {2020})}\BibitemShut {NoStop}%
\bibitem [{\citenamefont {Sau}\ \emph {et~al.}(2021)\citenamefont {Sau},
  \citenamefont {Ikeshoji}, \citenamefont {Takagi}, \citenamefont {Orimo},
  \citenamefont {Errandonea}, \citenamefont {Chu},\ and\ \citenamefont
  {Cazorla}}]{Sau_ScientificReports2021}%
  \BibitemOpen
  \bibfield  {author} {\bibinfo {author} {\bibfnamefont {Kartik}\ \bibnamefont
  {Sau}}, \bibinfo {author} {\bibfnamefont {Tamio}\ \bibnamefont {Ikeshoji}},
  \bibinfo {author} {\bibfnamefont {Shigeyuki}\ \bibnamefont {Takagi}},
  \bibinfo {author} {\bibfnamefont {Shin-ichi}\ \bibnamefont {Orimo}}, \bibinfo
  {author} {\bibfnamefont {Daniel}\ \bibnamefont {Errandonea}}, \bibinfo
  {author} {\bibfnamefont {Dewei}\ \bibnamefont {Chu}}, \ and\ \bibinfo
  {author} {\bibfnamefont {Claudio}\ \bibnamefont {Cazorla}},\ }\bibfield
  {title} {\enquote {\bibinfo {title} {Colossal barocaloric effects in the
  complex hydride {{Li}}$_{2}${{B}}$_{12}${{H}}$_{12}$},}\ }\href {\doibase
  10.1038/s41598-021-91123-4} {\bibfield  {journal} {\bibinfo  {journal}
  {Scientific Reports}\ }\textbf {\bibinfo {volume} {11}},\ \bibinfo {pages}
  {11915} (\bibinfo {year} {2021})}\BibitemShut {NoStop}%
\bibitem [{\citenamefont {Ren}\ \emph {et~al.}(2022)\citenamefont {Ren},
  \citenamefont {Qi}, \citenamefont {Yu}, \citenamefont {Zhang}, \citenamefont
  {Song}, \citenamefont {Song}, \citenamefont {Yuan}, \citenamefont {Wang},
  \citenamefont {Ren}, \citenamefont {Zhang}, \citenamefont {Tong},\ and\
  \citenamefont {Li}}]{LibingNC2022}%
  \BibitemOpen
  \bibfield  {author} {\bibinfo {author} {\bibfnamefont {Qingyong}\
  \bibnamefont {Ren}}, \bibinfo {author} {\bibfnamefont {Ji}~\bibnamefont
  {Qi}}, \bibinfo {author} {\bibfnamefont {Dehong}\ \bibnamefont {Yu}},
  \bibinfo {author} {\bibfnamefont {Zhe}\ \bibnamefont {Zhang}}, \bibinfo
  {author} {\bibfnamefont {Ruiqi}\ \bibnamefont {Song}}, \bibinfo {author}
  {\bibfnamefont {Wenli}\ \bibnamefont {Song}}, \bibinfo {author}
  {\bibfnamefont {Bao}\ \bibnamefont {Yuan}}, \bibinfo {author} {\bibfnamefont
  {Tianhao}\ \bibnamefont {Wang}}, \bibinfo {author} {\bibfnamefont {Weijun}\
  \bibnamefont {Ren}}, \bibinfo {author} {\bibfnamefont {Zhidong}\ \bibnamefont
  {Zhang}}, \bibinfo {author} {\bibfnamefont {Xin}\ \bibnamefont {Tong}}, \
  and\ \bibinfo {author} {\bibfnamefont {Bing}\ \bibnamefont {Li}},\ }\bibfield
   {title} {\enquote {\bibinfo {title} {Ultrasensitive barocaloric material for
  room-temperature solid-state refrigeration},}\ }\href {\doibase
  10.1038/s41467-022-29997-9} {\bibfield  {journal} {\bibinfo  {journal}
  {Nature Communications}\ }\textbf {\bibinfo {volume} {13}},\ \bibinfo {pages}
  {2293} (\bibinfo {year} {2022})}\BibitemShut {NoStop}%
\bibitem [{\citenamefont {Salvatori}\ \emph {et~al.}(2022)\citenamefont
  {Salvatori}, \citenamefont {Negrier}, \citenamefont {Aznar}, \citenamefont
  {Barrio}, \citenamefont {Tamarit},\ and\ \citenamefont
  {Lloveras}}]{Salvatori2022}%
  \BibitemOpen
  \bibfield  {author} {\bibinfo {author} {\bibfnamefont {Alejandro}\
  \bibnamefont {Salvatori}}, \bibinfo {author} {\bibfnamefont {Philippe}\
  \bibnamefont {Negrier}}, \bibinfo {author} {\bibfnamefont {Araceli}\
  \bibnamefont {Aznar}}, \bibinfo {author} {\bibfnamefont {Mar{\'i}a}\
  \bibnamefont {Barrio}}, \bibinfo {author} {\bibfnamefont {Josep~Llu{\'i}s}\
  \bibnamefont {Tamarit}}, \ and\ \bibinfo {author} {\bibfnamefont {Pol}\
  \bibnamefont {Lloveras}},\ }\bibfield  {title} {\enquote {\bibinfo {title}
  {Colossal barocaloric effects in adamantane derivatives for thermal
  management},}\ }\href {\doibase 10.1063/5.0127667} {\bibfield  {journal}
  {\bibinfo  {journal} {APL Materials}\ }\textbf {\bibinfo {volume} {10}},\
  \bibinfo {pages} {111117} (\bibinfo {year} {2022})}\BibitemShut {NoStop}%
\bibitem [{\citenamefont {Zhang}\ \emph {et~al.}(2022)\citenamefont {Zhang},
  \citenamefont {Song}, \citenamefont {Qi}, \citenamefont {Zhang},
  \citenamefont {Zhang}, \citenamefont {Yu}, \citenamefont {Li}, \citenamefont
  {Zhang},\ and\ \citenamefont {Li}}]{ZhangAFM2022}%
  \BibitemOpen
  \bibfield  {author} {\bibinfo {author} {\bibfnamefont {Kun}\ \bibnamefont
  {Zhang}}, \bibinfo {author} {\bibfnamefont {Ruiqi}\ \bibnamefont {Song}},
  \bibinfo {author} {\bibfnamefont {Ji}~\bibnamefont {Qi}}, \bibinfo {author}
  {\bibfnamefont {Zhe}\ \bibnamefont {Zhang}}, \bibinfo {author} {\bibfnamefont
  {Zhao}\ \bibnamefont {Zhang}}, \bibinfo {author} {\bibfnamefont {Chenyang}\
  \bibnamefont {Yu}}, \bibinfo {author} {\bibfnamefont {Kuo}\ \bibnamefont
  {Li}}, \bibinfo {author} {\bibfnamefont {Zhidong}\ \bibnamefont {Zhang}}, \
  and\ \bibinfo {author} {\bibfnamefont {Bing}\ \bibnamefont {Li}},\ }\bibfield
   {title} {\enquote {\bibinfo {title} {Colossal barocaloric effect in
  carboranes as a performance tradeoff},}\ }\href {\doibase
  10.1002/adfm.202112622} {\bibfield  {journal} {\bibinfo  {journal} {Advanced
  Functional Materials}\ }\textbf {\bibinfo {volume} {32}},\ \bibinfo {pages}
  {2112622} (\bibinfo {year} {2022})}\BibitemShut {NoStop}%
\bibitem [{\citenamefont {Zhang}\ \emph
  {et~al.}(2023{\natexlab{a}})\citenamefont {Zhang}, \citenamefont {Li},
  \citenamefont {Lin}, \citenamefont {Song}, \citenamefont {Yu}, \citenamefont
  {Wang}, \citenamefont {Wang}, \citenamefont {Kawaguchi}, \citenamefont
  {Zhang}, \citenamefont {Yu}, \citenamefont {Li}, \citenamefont {Chen},
  \citenamefont {He}, \citenamefont {Mole}, \citenamefont {Yuan}, \citenamefont
  {Ren}, \citenamefont {Qian}, \citenamefont {Cai}, \citenamefont {Yu},
  \citenamefont {Wang}, \citenamefont {Zhao}, \citenamefont {Tong},
  \citenamefont {Zhang},\ and\ \citenamefont {Li}}]{LibingSA2023}%
  \BibitemOpen
  \bibfield  {author} {\bibinfo {author} {\bibfnamefont {Zhe}\ \bibnamefont
  {Zhang}}, \bibinfo {author} {\bibfnamefont {Kuo}\ \bibnamefont {Li}},
  \bibinfo {author} {\bibfnamefont {Shangchao}\ \bibnamefont {Lin}}, \bibinfo
  {author} {\bibfnamefont {Ruiqi}\ \bibnamefont {Song}}, \bibinfo {author}
  {\bibfnamefont {Dehong}\ \bibnamefont {Yu}}, \bibinfo {author} {\bibfnamefont
  {Yida}\ \bibnamefont {Wang}}, \bibinfo {author} {\bibfnamefont {Jingfan}\
  \bibnamefont {Wang}}, \bibinfo {author} {\bibfnamefont {Shogo}\ \bibnamefont
  {Kawaguchi}}, \bibinfo {author} {\bibfnamefont {Zhao}\ \bibnamefont {Zhang}},
  \bibinfo {author} {\bibfnamefont {Chenyang}\ \bibnamefont {Yu}}, \bibinfo
  {author} {\bibfnamefont {Xiaodong}\ \bibnamefont {Li}}, \bibinfo {author}
  {\bibfnamefont {Jie}\ \bibnamefont {Chen}}, \bibinfo {author} {\bibfnamefont
  {Lunhua}\ \bibnamefont {He}}, \bibinfo {author} {\bibfnamefont {Richard}\
  \bibnamefont {Mole}}, \bibinfo {author} {\bibfnamefont {Bao}\ \bibnamefont
  {Yuan}}, \bibinfo {author} {\bibfnamefont {Qingyong}\ \bibnamefont {Ren}},
  \bibinfo {author} {\bibfnamefont {Kun}\ \bibnamefont {Qian}}, \bibinfo
  {author} {\bibfnamefont {Zhuangli}\ \bibnamefont {Cai}}, \bibinfo {author}
  {\bibfnamefont {Jingui}\ \bibnamefont {Yu}}, \bibinfo {author} {\bibfnamefont
  {Mingchao}\ \bibnamefont {Wang}}, \bibinfo {author} {\bibfnamefont
  {Changying}\ \bibnamefont {Zhao}}, \bibinfo {author} {\bibfnamefont {Xin}\
  \bibnamefont {Tong}}, \bibinfo {author} {\bibfnamefont {Zhidong}\
  \bibnamefont {Zhang}}, \ and\ \bibinfo {author} {\bibfnamefont {Bing}\
  \bibnamefont {Li}},\ }\bibfield  {title} {\enquote {\bibinfo {title} {Thermal
  batteries based on inverse barocaloric effects},}\ }\href {\doibase
  10.1126/sciadv.add0374} {\bibfield  {journal} {\bibinfo  {journal} {Science
  Advances}\ }\textbf {\bibinfo {volume} {9}},\ \bibinfo {pages} {eadd0374}
  (\bibinfo {year} {2023}{\natexlab{a}})}\BibitemShut {NoStop}%
\bibitem [{\citenamefont {Escorihuela-Sayalero}\ \emph
  {et~al.}(2024)\citenamefont {Escorihuela-Sayalero}, \citenamefont {Pardo},
  \citenamefont {Romanini}, \citenamefont {Obrecht}, \citenamefont
  {Loehl{\'e}}, \citenamefont {Lloveras}, \citenamefont {Tamarit},\ and\
  \citenamefont {Cazorla}}]{EscorihuelaSayalero2024}%
  \BibitemOpen
  \bibfield  {author} {\bibinfo {author} {\bibfnamefont {Carlos}\ \bibnamefont
  {Escorihuela-Sayalero}}, \bibinfo {author} {\bibfnamefont {Luis~Carlos}\
  \bibnamefont {Pardo}}, \bibinfo {author} {\bibfnamefont {Michela}\
  \bibnamefont {Romanini}}, \bibinfo {author} {\bibfnamefont {Nicolas}\
  \bibnamefont {Obrecht}}, \bibinfo {author} {\bibfnamefont {Sophie}\
  \bibnamefont {Loehl{\'e}}}, \bibinfo {author} {\bibfnamefont {Pol}\
  \bibnamefont {Lloveras}}, \bibinfo {author} {\bibfnamefont {Josep-Llu{\'i}s}\
  \bibnamefont {Tamarit}}, \ and\ \bibinfo {author} {\bibfnamefont {Claudio}\
  \bibnamefont {Cazorla}},\ }\bibfield  {title} {\enquote {\bibinfo {title}
  {Prediction and understanding of barocaloric effects in orientationally
  disordered materials from molecular dynamics simulations},}\ }\href {\doibase
  10.1038/s41524-024-01199-5} {\bibfield  {journal} {\bibinfo  {journal} {npj
  Computational Materials}\ }\textbf {\bibinfo {volume} {10}},\ \bibinfo
  {pages} {13} (\bibinfo {year} {2024})}\BibitemShut {NoStop}%
\bibitem [{\citenamefont {Piper}\ \emph {et~al.}(2025)\citenamefont {Piper},
  \citenamefont {Melag}, \citenamefont {Kar}, \citenamefont {Sourjah},
  \citenamefont {Xiao}, \citenamefont {May}, \citenamefont {Aguey-Zinsou},
  \citenamefont {MacFarlane},\ and\ \citenamefont
  {Pringle}}]{PiperScience2025}%
  \BibitemOpen
  \bibfield  {author} {\bibinfo {author} {\bibfnamefont {Samantha~L.}\
  \bibnamefont {Piper}}, \bibinfo {author} {\bibfnamefont {Leena}\ \bibnamefont
  {Melag}}, \bibinfo {author} {\bibfnamefont {Mega}\ \bibnamefont {Kar}},
  \bibinfo {author} {\bibfnamefont {Azra}\ \bibnamefont {Sourjah}}, \bibinfo
  {author} {\bibfnamefont {Xiong}\ \bibnamefont {Xiao}}, \bibinfo {author}
  {\bibfnamefont {Eric~F.}\ \bibnamefont {May}}, \bibinfo {author}
  {\bibfnamefont {Kondo-Francois}\ \bibnamefont {Aguey-Zinsou}}, \bibinfo
  {author} {\bibfnamefont {Douglas~R.}\ \bibnamefont {MacFarlane}}, \ and\
  \bibinfo {author} {\bibfnamefont {Jennifer~M.}\ \bibnamefont {Pringle}},\
  }\bibfield  {title} {\enquote {\bibinfo {title} {Organic ionic plastic
  crystals having colossal barocaloric effects for sustainable
  refrigeration},}\ }\href {\doibase 10.1126/science.adq8396} {\bibfield
  {journal} {\bibinfo  {journal} {Science}\ }\textbf {\bibinfo {volume}
  {387}},\ \bibinfo {pages} {56--62} (\bibinfo {year} {2025})}\BibitemShut
  {NoStop}%
\bibitem [{\citenamefont {Zhao}\ \emph {et~al.}(2025)\citenamefont {Zhao},
  \citenamefont {Zhang}, \citenamefont {Hattori}, \citenamefont {Wang},
  \citenamefont {Li}, \citenamefont {Jia}, \citenamefont {Li}, \citenamefont
  {Xue}, \citenamefont {Fan}, \citenamefont {Song}, \citenamefont {Zhu},
  \citenamefont {Liu}, \citenamefont {Chen}, \citenamefont {Zhang},\ and\
  \citenamefont {Li}}]{LibingNC2025}%
  \BibitemOpen
  \bibfield  {author} {\bibinfo {author} {\bibfnamefont {Xueting}\ \bibnamefont
  {Zhao}}, \bibinfo {author} {\bibfnamefont {Zhao}\ \bibnamefont {Zhang}},
  \bibinfo {author} {\bibfnamefont {Takanori}\ \bibnamefont {Hattori}},
  \bibinfo {author} {\bibfnamefont {Jiantao}\ \bibnamefont {Wang}}, \bibinfo
  {author} {\bibfnamefont {Lingli}\ \bibnamefont {Li}}, \bibinfo {author}
  {\bibfnamefont {Yating}\ \bibnamefont {Jia}}, \bibinfo {author}
  {\bibfnamefont {Wanwu}\ \bibnamefont {Li}}, \bibinfo {author} {\bibfnamefont
  {Jianing}\ \bibnamefont {Xue}}, \bibinfo {author} {\bibfnamefont {Xiaoyan}\
  \bibnamefont {Fan}}, \bibinfo {author} {\bibfnamefont {Ruiqi}\ \bibnamefont
  {Song}}, \bibinfo {author} {\bibfnamefont {Jinlong}\ \bibnamefont {Zhu}},
  \bibinfo {author} {\bibfnamefont {Peitao}\ \bibnamefont {Liu}}, \bibinfo
  {author} {\bibfnamefont {Xing-Qiu}\ \bibnamefont {Chen}}, \bibinfo {author}
  {\bibfnamefont {Zhidong}\ \bibnamefont {Zhang}}, \ and\ \bibinfo {author}
  {\bibfnamefont {Bing}\ \bibnamefont {Li}},\ }\bibfield  {title} {\enquote
  {\bibinfo {title} {All-temperature barocaloric effects at pressure-induced
  phase transitions},}\ }\href {\doibase 10.1038/s41467-025-63068-z} {\bibfield
   {journal} {\bibinfo  {journal} {Nature Communications}\ }\textbf {\bibinfo
  {volume} {16}},\ \bibinfo {pages} {7713} (\bibinfo {year}
  {2025})}\BibitemShut {NoStop}%
\bibitem [{\citenamefont {Zhang}\ \emph
  {et~al.}(2023{\natexlab{b}})\citenamefont {Zhang}, \citenamefont {Jiang},
  \citenamefont {Hattori}, \citenamefont {Xu}, \citenamefont {Li},
  \citenamefont {Yu}, \citenamefont {Zhang}, \citenamefont {Yu}, \citenamefont
  {Mole}, \citenamefont {Yano}, \citenamefont {Chen}, \citenamefont {He},
  \citenamefont {Wang}, \citenamefont {Wang}, \citenamefont {Li},\ and\
  \citenamefont {Zhang}}]{Zhang2023}%
  \BibitemOpen
  \bibfield  {author} {\bibinfo {author} {\bibfnamefont {Zhao}\ \bibnamefont
  {Zhang}}, \bibinfo {author} {\bibfnamefont {Xiaoming}\ \bibnamefont {Jiang}},
  \bibinfo {author} {\bibfnamefont {Takanori}\ \bibnamefont {Hattori}},
  \bibinfo {author} {\bibfnamefont {Xiong}\ \bibnamefont {Xu}}, \bibinfo
  {author} {\bibfnamefont {Min}\ \bibnamefont {Li}}, \bibinfo {author}
  {\bibfnamefont {Chenyang}\ \bibnamefont {Yu}}, \bibinfo {author}
  {\bibfnamefont {Zhe}\ \bibnamefont {Zhang}}, \bibinfo {author} {\bibfnamefont
  {Dehong}\ \bibnamefont {Yu}}, \bibinfo {author} {\bibfnamefont {Richard}\
  \bibnamefont {Mole}}, \bibinfo {author} {\bibfnamefont {Shin-ichiro}\
  \bibnamefont {Yano}}, \bibinfo {author} {\bibfnamefont {Jie}\ \bibnamefont
  {Chen}}, \bibinfo {author} {\bibfnamefont {Lunhua}\ \bibnamefont {He}},
  \bibinfo {author} {\bibfnamefont {Chin-Wei}\ \bibnamefont {Wang}}, \bibinfo
  {author} {\bibfnamefont {Hui}\ \bibnamefont {Wang}}, \bibinfo {author}
  {\bibfnamefont {Bing}\ \bibnamefont {Li}}, \ and\ \bibinfo {author}
  {\bibfnamefont {Zhidong}\ \bibnamefont {Zhang}},\ }\bibfield  {title}
  {\enquote {\bibinfo {title} {A colossal barocaloric effect induced by the
  creation of a high-pressure phase},}\ }\href {\doibase 10.1039/D2MH00905F}
  {\bibfield  {journal} {\bibinfo  {journal} {Materials Horizons}\ }\textbf
  {\bibinfo {volume} {10}},\ \bibinfo {pages} {977--982} (\bibinfo {year}
  {2023}{\natexlab{b}})}\BibitemShut {NoStop}%
\bibitem [{\citenamefont {Kresse}\ and\ \citenamefont
  {Furthm\"uller}(1996)}]{PhysRevB.54.11169}%
  \BibitemOpen
  \bibfield  {author} {\bibinfo {author} {\bibfnamefont {G.}~\bibnamefont
  {Kresse}}\ and\ \bibinfo {author} {\bibfnamefont {J.}~\bibnamefont
  {Furthm\"uller}},\ }\bibfield  {title} {\enquote {\bibinfo {title} {Efficient
  iterative schemes for ab initio total-energy calculations using a plane-wave
  basis set},}\ }\href {\doibase 10.1103/PhysRevB.54.11169} {\bibfield
  {journal} {\bibinfo  {journal} {Phys. Rev. B}\ }\textbf {\bibinfo {volume}
  {54}},\ \bibinfo {pages} {11169--11186} (\bibinfo {year} {1996})}\BibitemShut
  {NoStop}%
\bibitem [{\citenamefont {Bl\"ochl}(1994)}]{PhysRevB.50.17953}%
  \BibitemOpen
  \bibfield  {author} {\bibinfo {author} {\bibfnamefont {P.~E.}\ \bibnamefont
  {Bl\"ochl}},\ }\bibfield  {title} {\enquote {\bibinfo {title} {Projector
  augmented-wave method},}\ }\href {\doibase 10.1103/PhysRevB.50.17953}
  {\bibfield  {journal} {\bibinfo  {journal} {Phys. Rev. B}\ }\textbf {\bibinfo
  {volume} {50}},\ \bibinfo {pages} {17953--17979} (\bibinfo {year}
  {1994})}\BibitemShut {NoStop}%
\bibitem [{\citenamefont {Kresse}\ and\ \citenamefont
  {Joubert}(1999)}]{PhysRevB.59.1758}%
  \BibitemOpen
  \bibfield  {author} {\bibinfo {author} {\bibfnamefont {G.}~\bibnamefont
  {Kresse}}\ and\ \bibinfo {author} {\bibfnamefont {D.}~\bibnamefont
  {Joubert}},\ }\bibfield  {title} {\enquote {\bibinfo {title} {From ultrasoft
  pseudopotentials to the projector augmented-wave method},}\ }\href {\doibase
  10.1103/PhysRevB.59.1758} {\bibfield  {journal} {\bibinfo  {journal} {Phys.
  Rev. B}\ }\textbf {\bibinfo {volume} {59}},\ \bibinfo {pages} {1758--1775}
  (\bibinfo {year} {1999})}\BibitemShut {NoStop}%
\bibitem [{\citenamefont {Perdew}\ \emph {et~al.}(2008)\citenamefont {Perdew},
  \citenamefont {Ruzsinszky}, \citenamefont {Csonka}, \citenamefont {Vydrov},
  \citenamefont {Scuseria}, \citenamefont {Constantin}, \citenamefont {Zhou},\
  and\ \citenamefont {Burke}}]{PhysRevLett.100.136406}%
  \BibitemOpen
  \bibfield  {author} {\bibinfo {author} {\bibfnamefont {John~P.}\ \bibnamefont
  {Perdew}}, \bibinfo {author} {\bibfnamefont {Adrienn}\ \bibnamefont
  {Ruzsinszky}}, \bibinfo {author} {\bibfnamefont {G\'abor~I.}\ \bibnamefont
  {Csonka}}, \bibinfo {author} {\bibfnamefont {Oleg~A.}\ \bibnamefont
  {Vydrov}}, \bibinfo {author} {\bibfnamefont {Gustavo~E.}\ \bibnamefont
  {Scuseria}}, \bibinfo {author} {\bibfnamefont {Lucian~A.}\ \bibnamefont
  {Constantin}}, \bibinfo {author} {\bibfnamefont {Xiaolan}\ \bibnamefont
  {Zhou}}, \ and\ \bibinfo {author} {\bibfnamefont {Kieron}\ \bibnamefont
  {Burke}},\ }\bibfield  {title} {\enquote {\bibinfo {title} {Restoring the
  density-gradient expansion for exchange in solids and surfaces},}\ }\href
  {\doibase 10.1103/PhysRevLett.100.136406} {\bibfield  {journal} {\bibinfo
  {journal} {Phys. Rev. Lett.}\ }\textbf {\bibinfo {volume} {100}},\ \bibinfo
  {pages} {136406} (\bibinfo {year} {2008})}\BibitemShut {NoStop}%
\bibitem [{\citenamefont {Krukau}\ \emph {et~al.}(2006)\citenamefont {Krukau},
  \citenamefont {Vydrov}, \citenamefont {Izmaylov},\ and\ \citenamefont
  {Scuseria}}]{HSE06}%
  \BibitemOpen
  \bibfield  {author} {\bibinfo {author} {\bibfnamefont {Aliaksandr~V.}\
  \bibnamefont {Krukau}}, \bibinfo {author} {\bibfnamefont {Oleg~A.}\
  \bibnamefont {Vydrov}}, \bibinfo {author} {\bibfnamefont {Artur~F.}\
  \bibnamefont {Izmaylov}}, \ and\ \bibinfo {author} {\bibfnamefont
  {Gustavo~E.}\ \bibnamefont {Scuseria}},\ }\bibfield  {title} {\enquote
  {\bibinfo {title} {Influence of the exchange screening parameter on the
  performance of screened hybrid functionals},}\ }\href {\doibase
  10.1063/1.2404663} {\bibfield  {journal} {\bibinfo  {journal} {The Journal of
  Chemical Physics}\ }\textbf {\bibinfo {volume} {125}},\ \bibinfo {pages}
  {224106} (\bibinfo {year} {2006})}\BibitemShut {NoStop}%
\bibitem [{\citenamefont {Togo}\ and\ \citenamefont
  {Tanaka}(2015)}]{Togo2015-pho}%
  \BibitemOpen
  \bibfield  {author} {\bibinfo {author} {\bibfnamefont {Atsushi}\ \bibnamefont
  {Togo}}\ and\ \bibinfo {author} {\bibfnamefont {Isao}\ \bibnamefont
  {Tanaka}},\ }\bibfield  {title} {\enquote {\bibinfo {title} {First principles
  phonon calculations in materials science},}\ }\href {\doibase
  https://doi.org/10.1016/j.scriptamat.2015.07.021} {\bibfield  {journal}
  {\bibinfo  {journal} {Scripta Materialia}\ }\textbf {\bibinfo {volume}
  {108}},\ \bibinfo {pages} {1--5} (\bibinfo {year} {2015})}\BibitemShut
  {NoStop}%
\bibitem [{\citenamefont {Qian}\ \emph {et~al.}(2013)\citenamefont {Qian},
  \citenamefont {Dong}, \citenamefont {Zhou}, \citenamefont {Tian},
  \citenamefont {Oganov},\ and\ \citenamefont {Wang}}]{QIAN20132111}%
  \BibitemOpen
  \bibfield  {author} {\bibinfo {author} {\bibfnamefont {Guang-Rui}\
  \bibnamefont {Qian}}, \bibinfo {author} {\bibfnamefont {Xiao}\ \bibnamefont
  {Dong}}, \bibinfo {author} {\bibfnamefont {Xiang-Feng}\ \bibnamefont {Zhou}},
  \bibinfo {author} {\bibfnamefont {Yongjun}\ \bibnamefont {Tian}}, \bibinfo
  {author} {\bibfnamefont {Artem~R.}\ \bibnamefont {Oganov}}, \ and\ \bibinfo
  {author} {\bibfnamefont {Hui-Tian}\ \bibnamefont {Wang}},\ }\bibfield
  {title} {\enquote {\bibinfo {title} {Variable cell nudged elastic band method
  for studying solid–solid structural phase transitions},}\ }\href {\doibase
  https://doi.org/10.1016/j.cpc.2013.04.004} {\bibfield  {journal} {\bibinfo
  {journal} {Computer Physics Communications}\ }\textbf {\bibinfo {volume}
  {184}},\ \bibinfo {pages} {2111--2118} (\bibinfo {year} {2013})}\BibitemShut
  {NoStop}%
\bibitem [{\citenamefont {Liu}\ \emph {et~al.}(2023)\citenamefont {Liu},
  \citenamefont {Wang}, \citenamefont {Avargues}, \citenamefont {Verdi},
  \citenamefont {Singraber}, \citenamefont {Karsai}, \citenamefont {Chen},\
  and\ \citenamefont {Kresse}}]{PhysRevLett.130.078001}%
  \BibitemOpen
  \bibfield  {author} {\bibinfo {author} {\bibfnamefont {Peitao}\ \bibnamefont
  {Liu}}, \bibinfo {author} {\bibfnamefont {Jiantao}\ \bibnamefont {Wang}},
  \bibinfo {author} {\bibfnamefont {Noah}\ \bibnamefont {Avargues}}, \bibinfo
  {author} {\bibfnamefont {Carla}\ \bibnamefont {Verdi}}, \bibinfo {author}
  {\bibfnamefont {Andreas}\ \bibnamefont {Singraber}}, \bibinfo {author}
  {\bibfnamefont {Ferenc}\ \bibnamefont {Karsai}}, \bibinfo {author}
  {\bibfnamefont {Xing-Qiu}\ \bibnamefont {Chen}}, \ and\ \bibinfo {author}
  {\bibfnamefont {Georg}\ \bibnamefont {Kresse}},\ }\bibfield  {title}
  {\enquote {\bibinfo {title} {Combining machine learning and many-body
  calculations: coverage-dependent adsorption of {CO} on {Rh}(111)},}\ }\href
  {\doibase 10.1103/PhysRevLett.130.078001} {\bibfield  {journal} {\bibinfo
  {journal} {Physical Review Letters}\ }\textbf {\bibinfo {volume} {130}},\
  \bibinfo {pages} {078001} (\bibinfo {year} {2023})}\BibitemShut {NoStop}%
\bibitem [{\citenamefont {Liu}\ \emph {et~al.}(2024)\citenamefont {Liu},
  \citenamefont {Wang}, \citenamefont {Hu}, \citenamefont {Liu}, \citenamefont
  {Niu}, \citenamefont {Yan}, \citenamefont {Li}, \citenamefont {Yan},
  \citenamefont {Yang}, \citenamefont {Sun}, \citenamefont {Chen},
  \citenamefont {Kresse}, \citenamefont {Zuo},\ and\ \citenamefont
  {Chen}}]{Liu2024}%
  \BibitemOpen
  \bibfield  {author} {\bibinfo {author} {\bibfnamefont {Mingfeng}\
  \bibnamefont {Liu}}, \bibinfo {author} {\bibfnamefont {Jiantao}\ \bibnamefont
  {Wang}}, \bibinfo {author} {\bibfnamefont {Junwei}\ \bibnamefont {Hu}},
  \bibinfo {author} {\bibfnamefont {Peitao}\ \bibnamefont {Liu}}, \bibinfo
  {author} {\bibfnamefont {Haiyang}\ \bibnamefont {Niu}}, \bibinfo {author}
  {\bibfnamefont {Xuexi}\ \bibnamefont {Yan}}, \bibinfo {author} {\bibfnamefont
  {Jiangxu}\ \bibnamefont {Li}}, \bibinfo {author} {\bibfnamefont {Haile}\
  \bibnamefont {Yan}}, \bibinfo {author} {\bibfnamefont {Bo}~\bibnamefont
  {Yang}}, \bibinfo {author} {\bibfnamefont {Yan}\ \bibnamefont {Sun}},
  \bibinfo {author} {\bibfnamefont {Chunlin}\ \bibnamefont {Chen}}, \bibinfo
  {author} {\bibfnamefont {Georg}\ \bibnamefont {Kresse}}, \bibinfo {author}
  {\bibfnamefont {Liang}\ \bibnamefont {Zuo}}, \ and\ \bibinfo {author}
  {\bibfnamefont {Xing-Qiu}\ \bibnamefont {Chen}},\ }\bibfield  {title}
  {\enquote {\bibinfo {title} {Layer-by-layer phase transformation in
  {{Ti}}$_3${{O}}$_5$ revealed by machine-learning molecular dynamics
  simulations},}\ }\href {\doibase 10.1038/s41467-024-47422-1} {\bibfield
  {journal} {\bibinfo  {journal} {Nature Communications}\ }\textbf {\bibinfo
  {volume} {15}},\ \bibinfo {pages} {3079} (\bibinfo {year}
  {2024})}\BibitemShut {NoStop}%
\bibitem [{\citenamefont {Cao}\ \emph {et~al.}(2025)\citenamefont {Cao},
  \citenamefont {Wang}, \citenamefont {Liu}, \citenamefont {Liu}, \citenamefont
  {Ma}, \citenamefont {Franchini}, \citenamefont {Sun}, \citenamefont {Kresse},
  \citenamefont {Chen},\ and\ \citenamefont {Liu}}]{PhysRevLett.134.178001}%
  \BibitemOpen
  \bibfield  {author} {\bibinfo {author} {\bibfnamefont {Yu}~\bibnamefont
  {Cao}}, \bibinfo {author} {\bibfnamefont {Jiantao}\ \bibnamefont {Wang}},
  \bibinfo {author} {\bibfnamefont {Mingfeng}\ \bibnamefont {Liu}}, \bibinfo
  {author} {\bibfnamefont {Yan}\ \bibnamefont {Liu}}, \bibinfo {author}
  {\bibfnamefont {Hui}\ \bibnamefont {Ma}}, \bibinfo {author} {\bibfnamefont
  {Cesare}\ \bibnamefont {Franchini}}, \bibinfo {author} {\bibfnamefont {Yan}\
  \bibnamefont {Sun}}, \bibinfo {author} {\bibfnamefont {Georg}\ \bibnamefont
  {Kresse}}, \bibinfo {author} {\bibfnamefont {Xing-Qiu}\ \bibnamefont {Chen}},
  \ and\ \bibinfo {author} {\bibfnamefont {Peitao}\ \bibnamefont {Liu}},\
  }\bibfield  {title} {\enquote {\bibinfo {title} {Quantum delocalization
  enables water dissociation on {{Ru}}(0001)},}\ }\href {\doibase
  10.1103/PhysRevLett.134.178001} {\bibfield  {journal} {\bibinfo  {journal}
  {Phys. Rev. Lett.}\ }\textbf {\bibinfo {volume} {134}},\ \bibinfo {pages}
  {178001} (\bibinfo {year} {2025})}\BibitemShut {NoStop}%
\bibitem [{\citenamefont {Jinnouchi}\ \emph
  {et~al.}(2019{\natexlab{a}})\citenamefont {Jinnouchi}, \citenamefont
  {Lahnsteiner}, \citenamefont {Karsai}, \citenamefont {Kresse},\ and\
  \citenamefont {Bokdam}}]{JinnouchiPRL2019}%
  \BibitemOpen
  \bibfield  {author} {\bibinfo {author} {\bibfnamefont {Ryosuke}\ \bibnamefont
  {Jinnouchi}}, \bibinfo {author} {\bibfnamefont {Jonathan}\ \bibnamefont
  {Lahnsteiner}}, \bibinfo {author} {\bibfnamefont {Ferenc}\ \bibnamefont
  {Karsai}}, \bibinfo {author} {\bibfnamefont {Georg}\ \bibnamefont {Kresse}},
  \ and\ \bibinfo {author} {\bibfnamefont {Menno}\ \bibnamefont {Bokdam}},\
  }\bibfield  {title} {\enquote {\bibinfo {title} {Phase transitions of hybrid
  perovskites simulated by machine-learning force fields trained on the fly
  with bayesian inference},}\ }\href {\doibase 10.1103/PhysRevLett.122.225701}
  {\bibfield  {journal} {\bibinfo  {journal} {Physical Review Letters}\
  }\textbf {\bibinfo {volume} {122}},\ \bibinfo {pages} {225701} (\bibinfo
  {year} {2019}{\natexlab{a}})}\BibitemShut {NoStop}%
\bibitem [{\citenamefont {Jinnouchi}\ \emph
  {et~al.}(2019{\natexlab{b}})\citenamefont {Jinnouchi}, \citenamefont
  {Karsai},\ and\ \citenamefont {Kresse}}]{JinnouchiPRB2019}%
  \BibitemOpen
  \bibfield  {author} {\bibinfo {author} {\bibfnamefont {Ryosuke}\ \bibnamefont
  {Jinnouchi}}, \bibinfo {author} {\bibfnamefont {Ferenc}\ \bibnamefont
  {Karsai}}, \ and\ \bibinfo {author} {\bibfnamefont {Georg}\ \bibnamefont
  {Kresse}},\ }\bibfield  {title} {\enquote {\bibinfo {title} {On-the-fly
  machine learning force field generation: application to melting points},}\
  }\href {\doibase 10.1103/PhysRevB.100.014105} {\bibfield  {journal} {\bibinfo
   {journal} {Physical Review B}\ }\textbf {\bibinfo {volume} {100}},\ \bibinfo
  {pages} {014105} (\bibinfo {year} {2019}{\natexlab{b}})}\BibitemShut
  {NoStop}%
\bibitem [{\citenamefont {Allen}\ and\ \citenamefont
  {Schmid}(2007)}]{allenThermostatMolecularDynamics2007}%
  \BibitemOpen
  \bibfield  {author} {\bibinfo {author} {\bibfnamefont {Michael~P.}\
  \bibnamefont {Allen}}\ and\ \bibinfo {author} {\bibfnamefont {Friederike}\
  \bibnamefont {Schmid}},\ }\bibfield  {title} {\enquote {\bibinfo {title} {A
  thermostat for molecular dynamics of complex fluids},}\ }\href {\doibase
  10.1080/08927020601052856} {\bibfield  {journal} {\bibinfo  {journal}
  {Molecular Simulation}\ }\textbf {\bibinfo {volume} {33}},\ \bibinfo {pages}
  {21--26} (\bibinfo {year} {2007})}\BibitemShut {NoStop}%
\bibitem [{\citenamefont {Parrinello}\ and\ \citenamefont
  {Rahman}(1980)}]{parrinelloCrystalStructurePair1980}%
  \BibitemOpen
  \bibfield  {author} {\bibinfo {author} {\bibfnamefont {M.}~\bibnamefont
  {Parrinello}}\ and\ \bibinfo {author} {\bibfnamefont {A.}~\bibnamefont
  {Rahman}},\ }\bibfield  {title} {\enquote {\bibinfo {title} {Crystal
  structure and pair potentials: {{A}} molecular-dynamics study},}\ }\href
  {\doibase 10.1103/PhysRevLett.45.1196} {\bibfield  {journal} {\bibinfo
  {journal} {Physical Review Letters}\ }\textbf {\bibinfo {volume} {45}},\
  \bibinfo {pages} {1196--1199} (\bibinfo {year} {1980})}\BibitemShut {NoStop}%
\bibitem [{\citenamefont {Parrinello}\ and\ \citenamefont
  {Rahman}(1981)}]{parrinelloPolymorphicTransitionsSingle1981}%
  \BibitemOpen
  \bibfield  {author} {\bibinfo {author} {\bibfnamefont {M.}~\bibnamefont
  {Parrinello}}\ and\ \bibinfo {author} {\bibfnamefont {A.}~\bibnamefont
  {Rahman}},\ }\bibfield  {title} {\enquote {\bibinfo {title} {Polymorphic
  transitions in single crystals: {{A}} new molecular dynamics method},}\
  }\href {\doibase 10.1063/1.328693} {\bibfield  {journal} {\bibinfo  {journal}
  {Journal of Applied Physics}\ }\textbf {\bibinfo {volume} {52}},\ \bibinfo
  {pages} {7182--7190} (\bibinfo {year} {1981})}\BibitemShut {NoStop}%
\bibitem [{\citenamefont {Shapeev}(2016)}]{Alexander2016}%
  \BibitemOpen
  \bibfield  {author} {\bibinfo {author} {\bibfnamefont {Alexander~V.}\
  \bibnamefont {Shapeev}},\ }\bibfield  {title} {\enquote {\bibinfo {title}
  {Moment tensor potentials: a class of systematically improvable interatomic
  potentials},}\ }\href {\doibase 10.1137/15M1054183} {\bibfield  {journal}
  {\bibinfo  {journal} {Multiscale Modeling $\&$ Simulation}\ }\textbf
  {\bibinfo {volume} {14}},\ \bibinfo {pages} {1153--1173} (\bibinfo {year}
  {2016})}\BibitemShut {NoStop}%
\bibitem [{\citenamefont {Podryabinkin}\ and\ \citenamefont
  {Shapeev}(2017)}]{mtpal1}%
  \BibitemOpen
  \bibfield  {author} {\bibinfo {author} {\bibfnamefont {Evgeny~V.}\
  \bibnamefont {Podryabinkin}}\ and\ \bibinfo {author} {\bibfnamefont
  {Alexander~V.}\ \bibnamefont {Shapeev}},\ }\bibfield  {title} {\enquote
  {\bibinfo {title} {Active learning of linearly parametrized interatomic
  potentials},}\ }\href {\doibase 10.1016/j.commatsci.2017.08.031} {\bibfield
  {journal} {\bibinfo  {journal} {Computational Materials Science}\ }\textbf
  {\bibinfo {volume} {140}},\ \bibinfo {pages} {171--180} (\bibinfo {year}
  {2017})}\BibitemShut {NoStop}%
\bibitem [{\citenamefont {Zuo}\ \emph {et~al.}(2020)\citenamefont {Zuo},
  \citenamefont {Chen}, \citenamefont {Li}, \citenamefont {Deng}, \citenamefont
  {Chen}, \citenamefont {Behler}, \citenamefont {Cs{\'a}nyi}, \citenamefont
  {Shapeev}, \citenamefont {Thompson}, \citenamefont {Wood},\ and\
  \citenamefont {Ong}}]{zuoPerformanceCostAssessment2020}%
  \BibitemOpen
  \bibfield  {author} {\bibinfo {author} {\bibfnamefont {Yunxing}\ \bibnamefont
  {Zuo}}, \bibinfo {author} {\bibfnamefont {Chi}\ \bibnamefont {Chen}},
  \bibinfo {author} {\bibfnamefont {Xiangguo}\ \bibnamefont {Li}}, \bibinfo
  {author} {\bibfnamefont {Zhi}\ \bibnamefont {Deng}}, \bibinfo {author}
  {\bibfnamefont {Yiming}\ \bibnamefont {Chen}}, \bibinfo {author}
  {\bibfnamefont {J{\"o}rg}\ \bibnamefont {Behler}}, \bibinfo {author}
  {\bibfnamefont {G{\'a}bor}\ \bibnamefont {Cs{\'a}nyi}}, \bibinfo {author}
  {\bibfnamefont {Alexander~V.}\ \bibnamefont {Shapeev}}, \bibinfo {author}
  {\bibfnamefont {Aidan~P.}\ \bibnamefont {Thompson}}, \bibinfo {author}
  {\bibfnamefont {Mitchell~A.}\ \bibnamefont {Wood}}, \ and\ \bibinfo {author}
  {\bibfnamefont {Shyue~Ping}\ \bibnamefont {Ong}},\ }\bibfield  {title}
  {\enquote {\bibinfo {title} {A performance and cost assessment of machine
  learning interatomic potentials},}\ }\href {\doibase
  10.1021/acs.jpca.9b08723} {\bibfield  {journal} {\bibinfo  {journal} {The
  Journal of Physical Chemistry A}\ }\textbf {\bibinfo {volume} {124}},\
  \bibinfo {pages} {731--745} (\bibinfo {year} {2020})}\BibitemShut {NoStop}%
\bibitem [{\citenamefont {Thompson}\ \emph {et~al.}(2022)\citenamefont
  {Thompson}, \citenamefont {Aktulga}, \citenamefont {Berger}, \citenamefont
  {Bolintineanu}, \citenamefont {Brown}, \citenamefont {Crozier}, \citenamefont
  {In~'T~Veld}, \citenamefont {Kohlmeyer}, \citenamefont {Moore}, \citenamefont
  {Nguyen}, \citenamefont {Shan}, \citenamefont {Stevens}, \citenamefont
  {Tranchida}, \citenamefont {Trott},\ and\ \citenamefont
  {Plimpton}}]{thompsonLAMMPSFlexibleSimulation2022}%
  \BibitemOpen
  \bibfield  {author} {\bibinfo {author} {\bibfnamefont {Aidan~P.}\
  \bibnamefont {Thompson}}, \bibinfo {author} {\bibfnamefont {H.~Metin}\
  \bibnamefont {Aktulga}}, \bibinfo {author} {\bibfnamefont {Richard}\
  \bibnamefont {Berger}}, \bibinfo {author} {\bibfnamefont {Dan~S.}\
  \bibnamefont {Bolintineanu}}, \bibinfo {author} {\bibfnamefont {W.~Michael}\
  \bibnamefont {Brown}}, \bibinfo {author} {\bibfnamefont {Paul~S.}\
  \bibnamefont {Crozier}}, \bibinfo {author} {\bibfnamefont {Pieter~J.}\
  \bibnamefont {In~'T~Veld}}, \bibinfo {author} {\bibfnamefont {Axel}\
  \bibnamefont {Kohlmeyer}}, \bibinfo {author} {\bibfnamefont {Stan~G.}\
  \bibnamefont {Moore}}, \bibinfo {author} {\bibfnamefont {Trung~Dac}\
  \bibnamefont {Nguyen}}, \bibinfo {author} {\bibfnamefont {Ray}\ \bibnamefont
  {Shan}}, \bibinfo {author} {\bibfnamefont {Mark~J.}\ \bibnamefont {Stevens}},
  \bibinfo {author} {\bibfnamefont {Julien}\ \bibnamefont {Tranchida}},
  \bibinfo {author} {\bibfnamefont {Christian}\ \bibnamefont {Trott}}, \ and\
  \bibinfo {author} {\bibfnamefont {Steven~J.}\ \bibnamefont {Plimpton}},\
  }\bibfield  {title} {\enquote {\bibinfo {title} {{{LAMMPS}} - a flexible
  simulation tool for particle-based materials modeling at the atomic, meso,
  and continuum scales},}\ }\href {\doibase 10.1016/j.cpc.2021.108171}
  {\bibfield  {journal} {\bibinfo  {journal} {Computer Physics Communications}\
  }\textbf {\bibinfo {volume} {271}},\ \bibinfo {pages} {108171} (\bibinfo
  {year} {2022})}\BibitemShut {NoStop}%
\bibitem [{\citenamefont {Novikov}\ \emph {et~al.}(2021)\citenamefont
  {Novikov}, \citenamefont {Gubaev}, \citenamefont {Podryabinkin},\ and\
  \citenamefont {Shapeev}}]{Novikov_2021}%
  \BibitemOpen
  \bibfield  {author} {\bibinfo {author} {\bibfnamefont {Ivan~S}\ \bibnamefont
  {Novikov}}, \bibinfo {author} {\bibfnamefont {Konstantin}\ \bibnamefont
  {Gubaev}}, \bibinfo {author} {\bibfnamefont {Evgeny~V}\ \bibnamefont
  {Podryabinkin}}, \ and\ \bibinfo {author} {\bibfnamefont {Alexander~V}\
  \bibnamefont {Shapeev}},\ }\bibfield  {title} {\enquote {\bibinfo {title}
  {The {MLIP} package: moment tensor potentials with {MPI} and active
  learning},}\ }\href {\doibase 10.1088/2632-2153/abc9fe} {\bibfield  {journal}
  {\bibinfo  {journal} {Machine Learning: Science and Technology}\ }\textbf
  {\bibinfo {volume} {2}},\ \bibinfo {pages} {025002} (\bibinfo {year}
  {2021})}\BibitemShut {NoStop}%
\bibitem [{\citenamefont {Wang}\ \emph {et~al.}(2025)\citenamefont {Wang},
  \citenamefont {Liu}, \citenamefont {Zhu}, \citenamefont {Liu}, \citenamefont
  {Ma}, \citenamefont {Chen}, \citenamefont {Sun},\ and\ \citenamefont
  {Chen}}]{wang2024}%
  \BibitemOpen
  \bibfield  {author} {\bibinfo {author} {\bibfnamefont {Jiantao}\ \bibnamefont
  {Wang}}, \bibinfo {author} {\bibfnamefont {Peitao}\ \bibnamefont {Liu}},
  \bibinfo {author} {\bibfnamefont {Heyu}\ \bibnamefont {Zhu}}, \bibinfo
  {author} {\bibfnamefont {Mingfeng}\ \bibnamefont {Liu}}, \bibinfo {author}
  {\bibfnamefont {Hui}\ \bibnamefont {Ma}}, \bibinfo {author} {\bibfnamefont
  {Yun}\ \bibnamefont {Chen}}, \bibinfo {author} {\bibfnamefont {Yan}\
  \bibnamefont {Sun}}, \ and\ \bibinfo {author} {\bibfnamefont {Xing-Qiu}\
  \bibnamefont {Chen}},\ }\bibfield  {title} {\enquote {\bibinfo {title}
  {Efficient moment tensor machine-learning interatomic potential for accurate
  description of defects in {{Ni}}-{{Al}} alloys},}\ }\href {\doibase
  10.1103/PhysRevMaterials.9.053805} {\bibfield  {journal} {\bibinfo  {journal}
  {Phys. Rev. Mater.}\ }\textbf {\bibinfo {volume} {9}},\ \bibinfo {pages}
  {053805} (\bibinfo {year} {2025})}\BibitemShut {NoStop}%
\bibitem [{\citenamefont {Cheng}\ \emph {et~al.}(2020)\citenamefont {Cheng},
  \citenamefont {Griffiths}, \citenamefont {Wengert}, \citenamefont {Kunkel},
  \citenamefont {Stenczel}, \citenamefont {Zhu}, \citenamefont {Deringer},
  \citenamefont {Bernstein}, \citenamefont {Margraf}, \citenamefont {Reuter},\
  and\ \citenamefont {Csanyi}}]{chengMappingMaterialsMolecules2020}%
  \BibitemOpen
  \bibfield  {author} {\bibinfo {author} {\bibfnamefont {Bingqing}\
  \bibnamefont {Cheng}}, \bibinfo {author} {\bibfnamefont {Ryan-Rhys}\
  \bibnamefont {Griffiths}}, \bibinfo {author} {\bibfnamefont {Simon}\
  \bibnamefont {Wengert}}, \bibinfo {author} {\bibfnamefont {Christian}\
  \bibnamefont {Kunkel}}, \bibinfo {author} {\bibfnamefont {Tamas}\
  \bibnamefont {Stenczel}}, \bibinfo {author} {\bibfnamefont {Bonan}\
  \bibnamefont {Zhu}}, \bibinfo {author} {\bibfnamefont {Volker~L.}\
  \bibnamefont {Deringer}}, \bibinfo {author} {\bibfnamefont {Noam}\
  \bibnamefont {Bernstein}}, \bibinfo {author} {\bibfnamefont {Johannes~T.}\
  \bibnamefont {Margraf}}, \bibinfo {author} {\bibfnamefont {Karsten}\
  \bibnamefont {Reuter}}, \ and\ \bibinfo {author} {\bibfnamefont {Gabor}\
  \bibnamefont {Csanyi}},\ }\bibfield  {title} {\enquote {\bibinfo {title}
  {Mapping materials and molecules},}\ }\href {\doibase
  10.1021/acs.accounts.0c00403} {\bibfield  {journal} {\bibinfo  {journal}
  {Accounts of Chemical Research}\ }\textbf {\bibinfo {volume} {53}},\ \bibinfo
  {pages} {1981--1991} (\bibinfo {year} {2020})}\BibitemShut {NoStop}%
\bibitem [{\citenamefont {Alexa}(2022)}]{10.1109/CVPR52688.2022.00811}%
  \BibitemOpen
  \bibfield  {author} {\bibinfo {author} {\bibfnamefont {Marc}\ \bibnamefont
  {Alexa}},\ }\bibfield  {title} {\enquote {\bibinfo {title} {Super-fibonacci
  spirals: Fast, low-discrepancy sampling of so(3)},}\ }in\ \href {\doibase
  10.1109/CVPR52688.2022.00811} {\emph {\bibinfo {booktitle} {2022 IEEE/CVF
  Conference on Computer Vision and Pattern Recognition (CVPR)}}}\ (\bibinfo
  {year} {2022})\ pp.\ \bibinfo {pages} {8281--8290}\BibitemShut {NoStop}%
\bibitem [{SM()}]{SM}%
  \BibitemOpen
  \href@noop {} {\emph {\bibinfo {title} {\rm See Supplemental Material for the
  details of crystallographic information of the four phases of KPF$_6$,
  simulated X-ray diffraction patterns, additional phonon dispersion
  relationships of the C and R phases at 0 K and 0 GPa, addtional energy-volume
  curves inclduding the C phase, phonon mode analysis of the R phase, and
  element-resolved electronic density of states.}}}\BibitemShut {Stop}%
\bibitem [{\citenamefont {Huber}\ \emph {et~al.}(1997)\citenamefont {Huber},
  \citenamefont {Krummeck}, \citenamefont {Baller}, \citenamefont {Kr{\"u}ger},
  \citenamefont {Knorr}, ,\ and\ \citenamefont {Hauss{\"u}hl}}]{Huber1997}%
  \BibitemOpen
  \bibfield  {author} {\bibinfo {author} {\bibfnamefont {P.}~\bibnamefont
  {Huber}}, \bibinfo {author} {\bibnamefont {Krummeck}}, \bibinfo {author}
  {\bibnamefont {Baller}}, \bibinfo {author} {\bibnamefont {Kr{\"u}ger}},
  \bibinfo {author} {\bibnamefont {Knorr}}, , \ and\ \bibinfo {author}
  {\bibfnamefont {S.}~\bibnamefont {Hauss{\"u}hl}},\ }\bibfield  {title}
  {\enquote {\bibinfo {title} {Phases and phase transitions of {{KPF}}$_6$},}\
  }\href {\doibase 10.1080/00150199708012847} {\bibfield  {journal} {\bibinfo
  {journal} {Ferroelectrics}\ }\textbf {\bibinfo {volume} {203}},\ \bibinfo
  {pages} {211--219} (\bibinfo {year} {1997})}\BibitemShut {NoStop}%
\bibitem [{\citenamefont {Glass}\ \emph {et~al.}(2006)\citenamefont {Glass},
  \citenamefont {Oganov},\ and\ \citenamefont {Hansen}}]{Glass2006}%
  \BibitemOpen
  \bibfield  {author} {\bibinfo {author} {\bibfnamefont {Colin~W.}\
  \bibnamefont {Glass}}, \bibinfo {author} {\bibfnamefont {Artem~R.}\
  \bibnamefont {Oganov}}, \ and\ \bibinfo {author} {\bibfnamefont {Nikolaus}\
  \bibnamefont {Hansen}},\ }\bibfield  {title} {\enquote {\bibinfo {title}
  {{{USPEX}}---{{E}}volutionary crystal structure prediction},}\ }\href
  {https://www.sciencedirect.com/science/article/pii/S0010465506002931}
  {\bibfield  {journal} {\bibinfo  {journal} {Computer Physics Communications}\
  }\textbf {\bibinfo {volume} {175}},\ \bibinfo {pages} {713--720} (\bibinfo
  {year} {2006})}\BibitemShut {NoStop}%
\bibitem [{\citenamefont {Grimme}\ \emph {et~al.}(2010)\citenamefont {Grimme},
  \citenamefont {Antony}, \citenamefont {Ehrlich},\ and\ \citenamefont
  {Krieg}}]{10.1063/1.3382344}%
  \BibitemOpen
  \bibfield  {author} {\bibinfo {author} {\bibfnamefont {Stefan}\ \bibnamefont
  {Grimme}}, \bibinfo {author} {\bibfnamefont {Jens}\ \bibnamefont {Antony}},
  \bibinfo {author} {\bibfnamefont {Stephan}\ \bibnamefont {Ehrlich}}, \ and\
  \bibinfo {author} {\bibfnamefont {Helge}\ \bibnamefont {Krieg}},\ }\bibfield
  {title} {\enquote {\bibinfo {title} {{A consistent and accurate ab initio
  parametrization of density functional dispersion correction (DFT-D) for the
  94 elements H-Pu}},}\ }\href {\doibase 10.1063/1.3382344} {\bibfield
  {journal} {\bibinfo  {journal} {The Journal of Chemical Physics}\ }\textbf
  {\bibinfo {volume} {132}},\ \bibinfo {pages} {154104} (\bibinfo {year}
  {2010})}\BibitemShut {NoStop}%
\bibitem [{\citenamefont {Sun}\ \emph {et~al.}(2015)\citenamefont {Sun},
  \citenamefont {Ruzsinszky},\ and\ \citenamefont
  {Perdew}}]{PhysRevLett.115.036402}%
  \BibitemOpen
  \bibfield  {author} {\bibinfo {author} {\bibfnamefont {Jianwei}\ \bibnamefont
  {Sun}}, \bibinfo {author} {\bibfnamefont {Adrienn}\ \bibnamefont
  {Ruzsinszky}}, \ and\ \bibinfo {author} {\bibfnamefont {John~P.}\
  \bibnamefont {Perdew}},\ }\bibfield  {title} {\enquote {\bibinfo {title}
  {Strongly constrained and appropriately normed semilocal density
  functional},}\ }\href {\doibase 10.1103/PhysRevLett.115.036402} {\bibfield
  {journal} {\bibinfo  {journal} {Phys. Rev. Lett.}\ }\textbf {\bibinfo
  {volume} {115}},\ \bibinfo {pages} {036402} (\bibinfo {year}
  {2015})}\BibitemShut {NoStop}%
\bibitem [{\citenamefont {Monacelli}\ \emph {et~al.}(2021)\citenamefont
  {Monacelli}, \citenamefont {Bianco}, \citenamefont {Cherubini}, \citenamefont
  {Calandra}, \citenamefont {Errea},\ and\ \citenamefont
  {Mauri}}]{monacelliStochasticSelfconsistentHarmonic2021}%
  \BibitemOpen
  \bibfield  {author} {\bibinfo {author} {\bibfnamefont {Lorenzo}\ \bibnamefont
  {Monacelli}}, \bibinfo {author} {\bibfnamefont {Raffaello}\ \bibnamefont
  {Bianco}}, \bibinfo {author} {\bibfnamefont {Marco}\ \bibnamefont
  {Cherubini}}, \bibinfo {author} {\bibfnamefont {Matteo}\ \bibnamefont
  {Calandra}}, \bibinfo {author} {\bibfnamefont {Ion}\ \bibnamefont {Errea}}, \
  and\ \bibinfo {author} {\bibfnamefont {Francesco}\ \bibnamefont {Mauri}},\
  }\bibfield  {title} {\enquote {\bibinfo {title} {The stochastic
  self-consistent harmonic approximation: Calculating vibrational properties of
  materials with full quantum and anharmonic effects},}\ }\href {\doibase
  10.1088/1361-648X/ac066b} {\bibfield  {journal} {\bibinfo  {journal} {J.
  Phys.: Condens. Matter}\ }\textbf {\bibinfo {volume} {33}},\ \bibinfo {pages}
  {363001} (\bibinfo {year} {2021})}\BibitemShut {NoStop}%
\bibitem [{\citenamefont {Wang}\ \emph {et~al.}(2024)\citenamefont {Wang},
  \citenamefont {Ye}, \citenamefont {Zhu},\ and\ \citenamefont
  {Li}}]{wang2024crystal}%
  \BibitemOpen
  \bibfield  {author} {\bibinfo {author} {\bibfnamefont {Fang-Cheng}\
  \bibnamefont {Wang}}, \bibinfo {author} {\bibfnamefont {Qi-Jun}\ \bibnamefont
  {Ye}}, \bibinfo {author} {\bibfnamefont {Yu-Cheng}\ \bibnamefont {Zhu}}, \
  and\ \bibinfo {author} {\bibfnamefont {Xin-Zheng}\ \bibnamefont {Li}},\
  }\bibfield  {title} {\enquote {\bibinfo {title} {Crystal-structure matches in
  solid-solid phase transitions},}\ }\href {\doibase
  10.1103/PhysRevLett.132.086101} {\bibfield  {journal} {\bibinfo  {journal}
  {Phys. Rev. Lett.}\ }\textbf {\bibinfo {volume} {132}},\ \bibinfo {pages}
  {086101} (\bibinfo {year} {2024})}\BibitemShut {NoStop}%
\end{thebibliography}%

\end{document}


\title{Supplemental Material to \\
``Atomistic mechanisms of phase transitions in all-temperature barocaloric material KPF$_6$"}

\author{Jiantao Wang}
\affiliation{Shenyang National Laboratory for Materials Science, Institute of Metal Research, Chinese Academy of Sciences, 110016 Shenyang, China}
\affiliation{School of Materials Science and Engineering, University of Science and Technology of China, 110016 Shenyang, China}

\author{Yi-Chi Zhang}
\affiliation{Shenyang National Laboratory for Materials Science, Institute of Metal Research, Chinese Academy of Sciences, 110016 Shenyang, China}
\affiliation{School of Materials Science and Engineering, University of Science and Technology of China, 110016 Shenyang, China}

\author{Yan Liu}
\affiliation{Shenyang National Laboratory for Materials Science, Institute of Metal Research, Chinese Academy of Sciences, 110016 Shenyang, China}
\affiliation{School of Materials Science and Engineering, University of Science and Technology of China, 110016 Shenyang, China}

\author{Hongkun Deng}
\affiliation{Shenyang National Laboratory for Materials Science, Institute of Metal Research, Chinese Academy of Sciences, 110016 Shenyang, China}
\affiliation{School of Materials Science and Engineering, University of Science and Technology of China, 110016 Shenyang, China}

\author{Mingfeng Liu}
\affiliation{Shenyang National Laboratory for Materials Science, Institute of Metal Research, Chinese Academy of Sciences, 110016 Shenyang, China}

\author{Yan Sun}
\affiliation{Shenyang National Laboratory for Materials Science, Institute of Metal Research, Chinese Academy of Sciences, 110016 Shenyang, China}

\author{Bing Li}
\affiliation{Shenyang National Laboratory for Materials Science, Institute of Metal Research, Chinese Academy of Sciences, 110016 Shenyang, China}

\author{Xing-Qiu Chen}
\email{xingqiu.chen@imr.ac.cn}
\affiliation{Shenyang National Laboratory for Materials Science, Institute of Metal Research, Chinese Academy of Sciences, 110016 Shenyang, China}

\author{Peitao Liu}
\email{ptliu@imr.ac.cn}
\affiliation{Shenyang National Laboratory for Materials Science, Institute of Metal Research, Chinese Academy of Sciences, 110016 Shenyang, China}

\maketitle

\begin{figure}
\begin{center}
\includegraphics[width=0.6\textwidth,trim = {0.0cm 0.0cm 0.0cm 0.0cm}, clip]{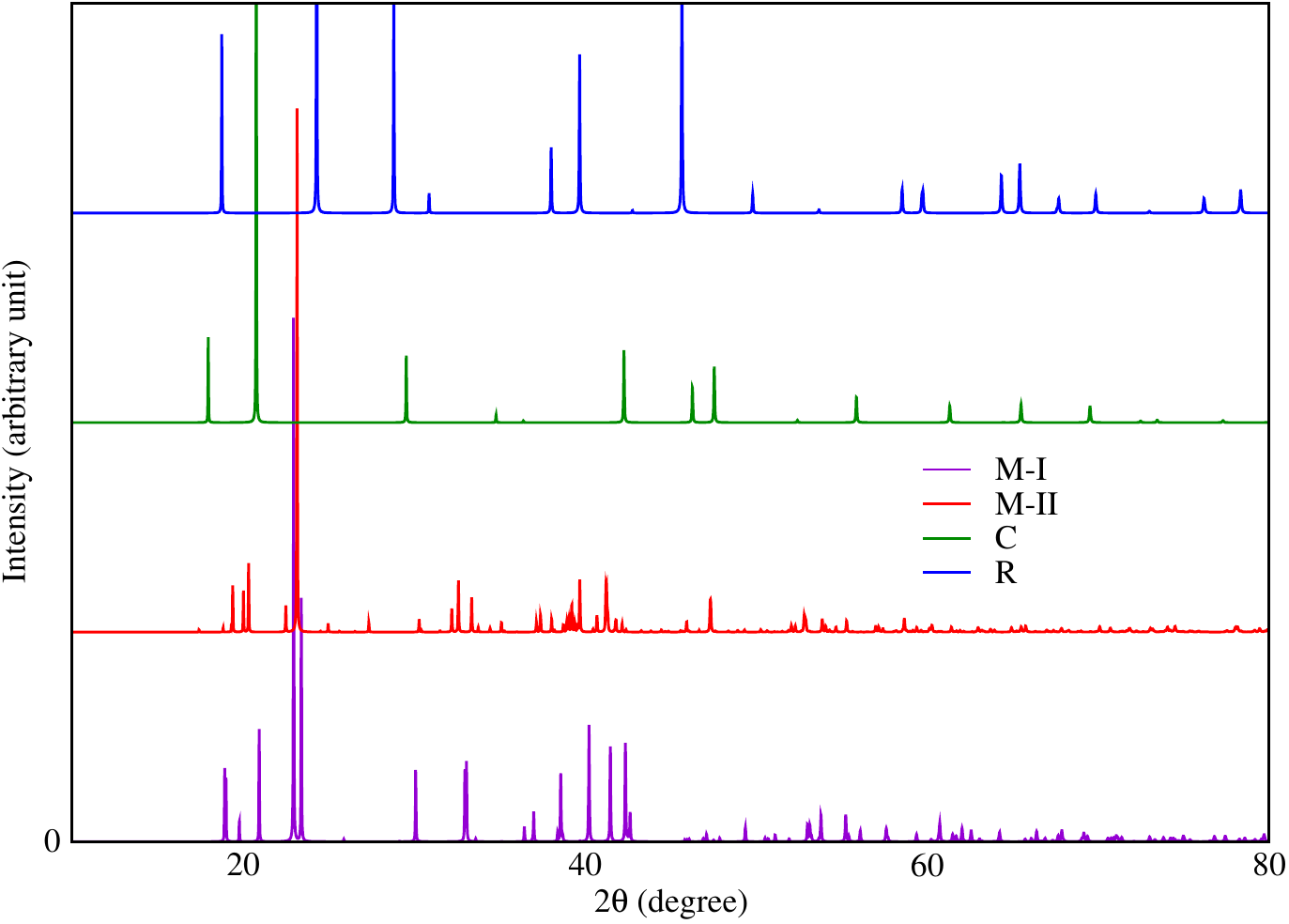}
\end{center}
\caption{Simulated X-ray diffraction patterns of the four phases of KPF$_6$ using the Cu $K_\alpha$ radiation.
The structures are obtained at 0 K and 0 GPa using the PBEsol functional.
}
\label{FigS1_XRD_PBEsol}
\end{figure}

\begin{figure}
\begin{center}
\includegraphics[width=0.6\textwidth,trim = {0.0cm 0.0cm 0.0cm 0.0cm}, clip]{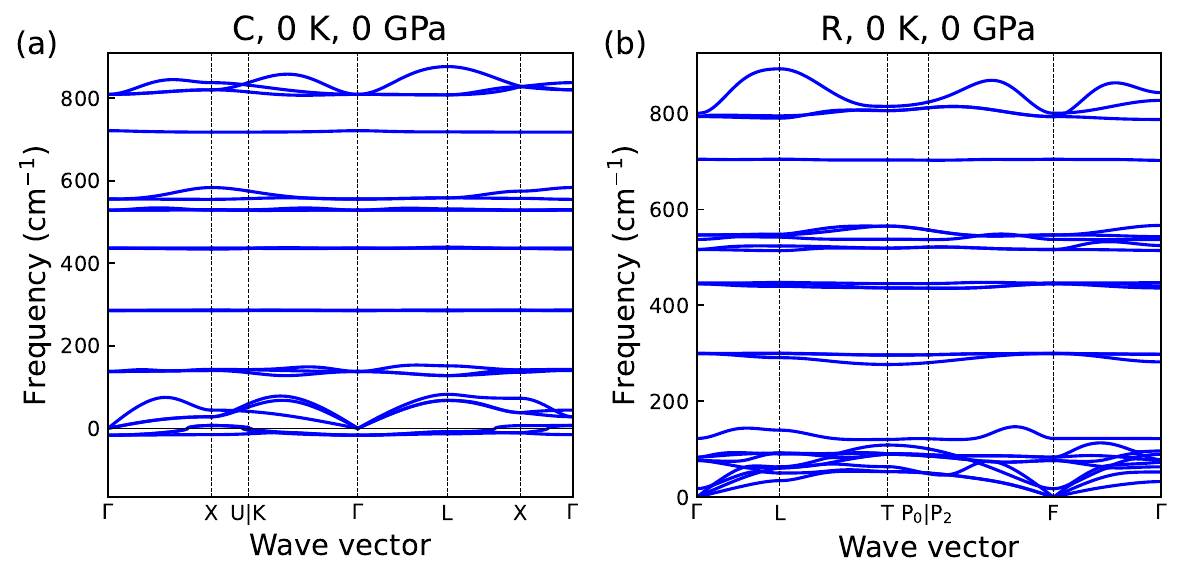}
\end{center}
\caption{Phonon dispersion relationships of (a) the C phase and (b) the R phase at 0 K and 0 GPa predicted using the PBEsol-derived MTP.
}
\label{FigS2_phonon}
\end{figure}

\begin{figure*}
\begin{center}
\includegraphics[width=0.9\textwidth,trim = {0.0cm 0.0cm 0.0cm 0.0cm}, clip]{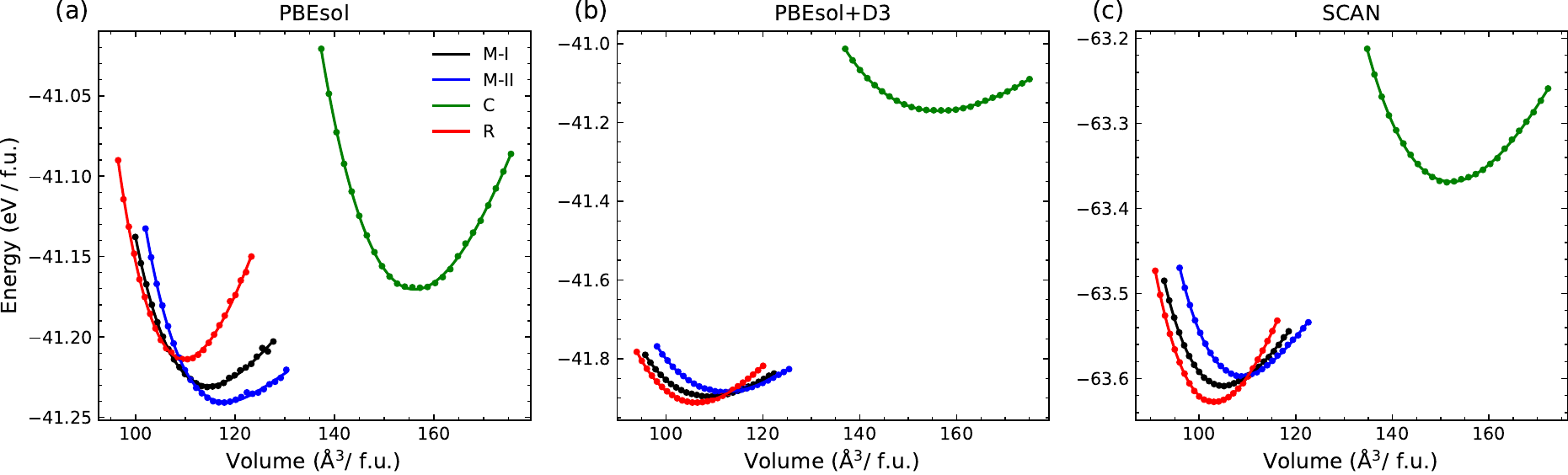}
\end{center}
\caption{Energy-volume curves for the four phases of KPF$_6$, calculated at 0 K using the (a) PBEsol, (b) PBEsol+D3, and (c) SCAN functionals.
}
\label{FigS3_E_vs_V}
\end{figure*}

\begin{figure}
\begin{center}
\includegraphics[width=0.7\textwidth,trim = {0.0cm 0.0cm 0.0cm 0.0cm}, clip]{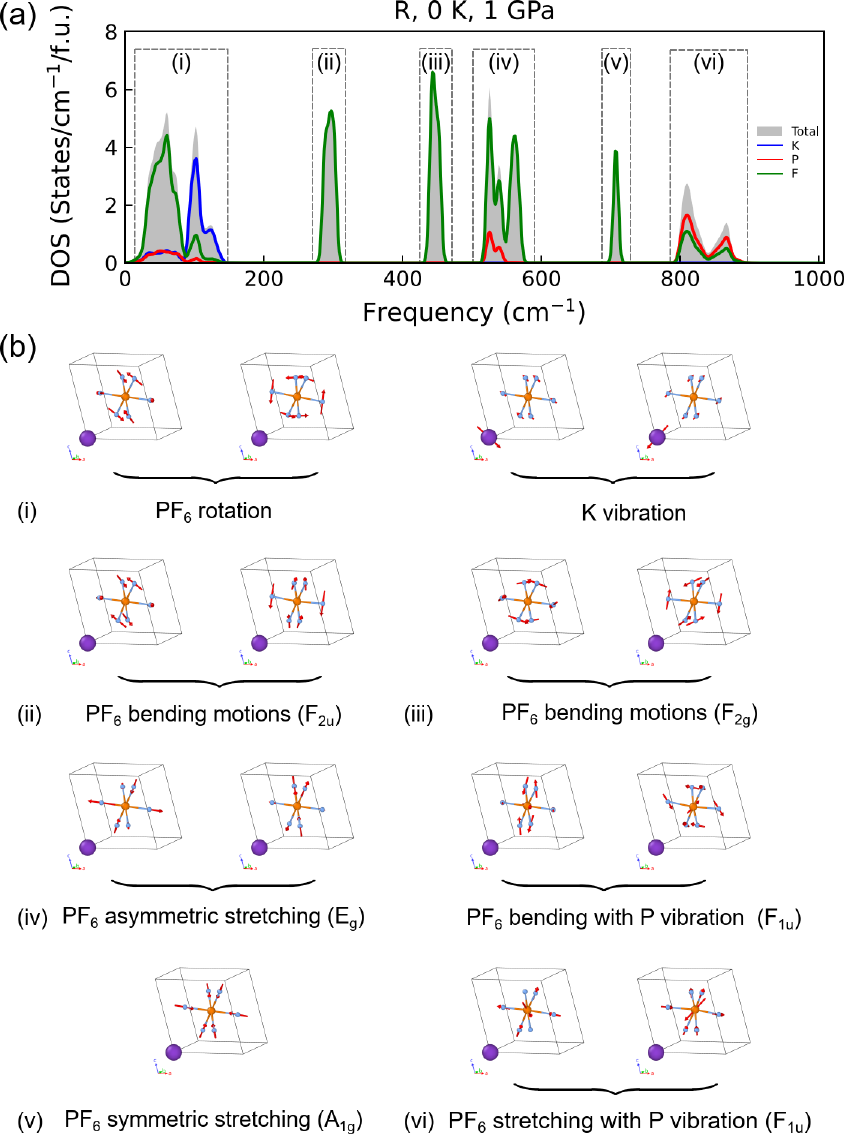}
\end{center}
\caption{The phonon mode analysis of the R phase of KPF$_6$, which is similar to those of the M-I, M-II, and C phases.
}
\label{FigS4_pdosR-mode}
\end{figure}

\begin{figure}
\begin{center}
\includegraphics[width=0.5\textwidth,trim = {0.0cm 0.0cm 0.0cm 0.0cm}, clip]{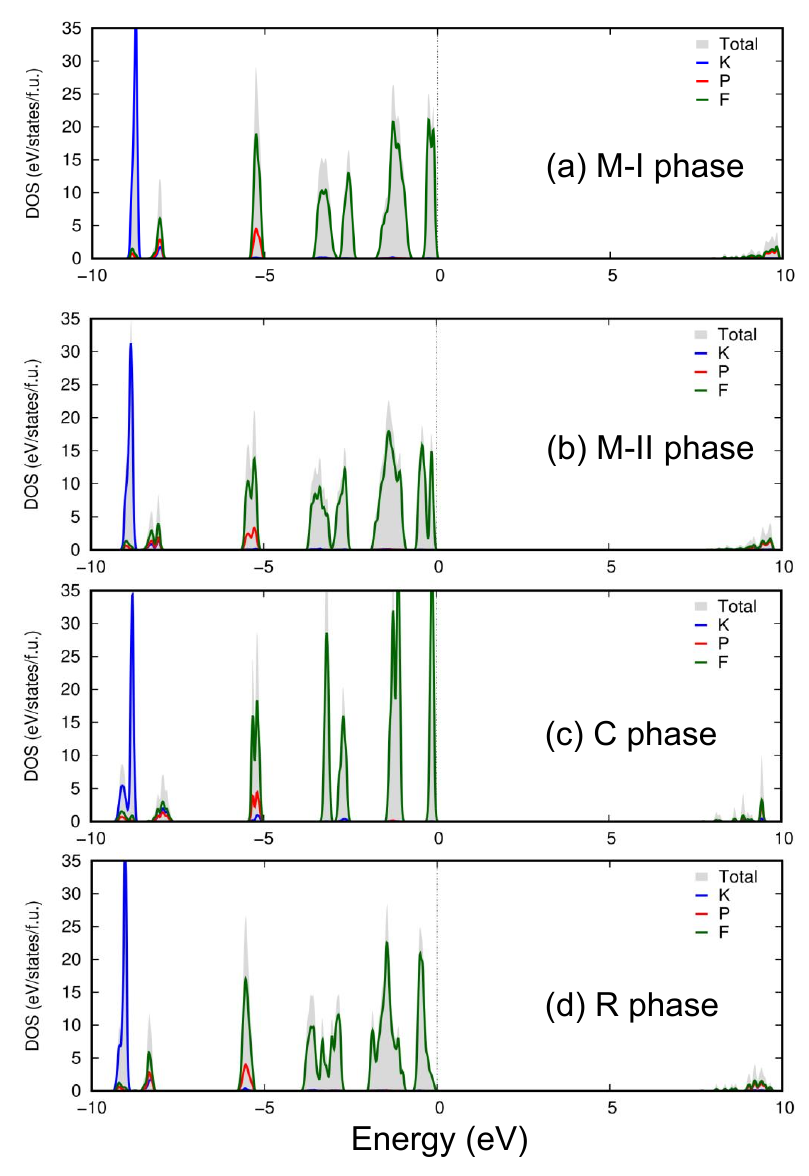}
\end{center}
\caption{Electronic density of states of the four phases of KPF$_6$ calculated using the HSE06 functional.
The structures are obtained at 0 K and 0 GPa using the PBEsol functional.
}
\label{FigS5_elec_dos}
\end{figure}

\clearpage
\newpage
\begin{table}
\caption{Crystallographic information of  the M-I phase of KPF$_6$ predicted by PBEsol at 0 K and 0 GPa.
}
\begin{ruledtabular}
\begin{tabular}{ccccc}
Space group & $a$  (\AA)  & $b$ (\AA)  &$c$ (\AA) & $\beta$ ($^{\rm o}$)  \\
\hline
$C2/c$ &9.655 & 5.106 & 9.617  & 104.00    \\
\hline
\multicolumn{5}{c}{Atomic positions} \\
Atoms & Site & $x$  & $y$ & $z$ \\
\hline
K &  4e  &  0.00000  &  0.33266  &  0.25000 \\
P &  4c  &  0.25000  &  0.25000  &  0.00000 \\
F &  8f  &  0.15534  &  0.49628  &  0.03271 \\
F &  8f  &  0.89330  &  0.12212  &  0.60295 \\
F &  8f  &  0.28025  &  0.40577  &  0.86241 \\
\end{tabular}
\end{ruledtabular}
\label{TableS1}
\end{table}

\begin{table}
\caption{Crystallographic information of  the M-II phase of KPF$_6$ predicted by PBEsol at 0 K and 0 GPa.
}
\begin{ruledtabular}
\begin{tabular}{ccccc}
Space group & $a$  (\AA)  & $b$ (\AA)  &$c$ (\AA) & $\beta$ ($^{\rm o}$)  \\
\hline
$Cc$ & 18.121 & 5.306 & 9.626  & 102.07    \\
\hline
\multicolumn{5}{c}{Atomic positions} \\
Atoms & Site & $x$  & $y$ & $z$ \\
\hline

K &  4a  &  0.12182   &  0.18750  &  0.15100 \\
K &  4a  &  0.38284   &  0.32796  &  0.86604 \\
P &  4a  &  0.75033   &  0.25663  &  0.00616 \\
P &  4a  &  -0.00103  &  0.32409  &  0.74145 \\
F &  4a  &  0.75331   &  0.14984  &  0.84841 \\
F &  4a  &  0.74752   &  0.36175  &  0.16394 \\
F &  4a  &  0.32880   &  0.39216  &  0.57190 \\
F &  4a  &  0.17176   &  0.09503  &  0.44031 \\
F &  4a  &  0.79802   &  0.49729  &  0.47802 \\
F &  4a  &  0.20225   &  0.48968  &  0.53343 \\
F &  4a  &  0.41395   &  0.16611  &  0.27250 \\
F &  4a  &  0.08444   &  0.31157  &  0.71173 \\
F &  4a  &  0.52657   &  0.33539  &  0.38779 \\
F &  4a  &  -0.02104  &  0.05918  &  0.65500 \\
F &  4a  &  0.47091   &  0.02132  &  0.09464 \\
F &  4a  &  0.51969   &  0.08771  &  0.82658 \\
\end{tabular}
\end{ruledtabular}
\label{TableS2}
\end{table}

\begin{table}
\caption{Crystallographic information of  the C phase of KPF$_6$ predicted by PBEsol at 0 K and 0 GPa.
}
\begin{ruledtabular}
\begin{tabular}{ccccc}
Space group & $a$  (\AA)  & $b$ (\AA)  &$c$ (\AA) & $\beta$ ($^{\rm o}$)  \\
\hline
$Fm\bar{3}m$ & 8.543 & 8.543 & 8.543  & 90.00    \\
\hline
\multicolumn{5}{c}{Atomic positions} \\
Atoms & Site & $x$  & $y$ & $z$ \\
\hline
K &  4a & 0.00000 & 0.00000 & 0.00000 \\
P &  4b & 0.50000 & 0.50000 & 0.50000 \\
F & 24e & 0.30948 & 0.00000 & 0.00000 \\
\end{tabular}
\end{ruledtabular}
\label{TableS3}
\end{table}

\begin{table}
\caption{Crystallographic information of  the R phase of KPF$_6$ predicted by PBEsol at 0 K and 0 GPa.
}
\begin{ruledtabular}
\begin{tabular}{ccccc}
Space group & $a$  (\AA)  & $b$ (\AA)  &$c$ (\AA) & $\gamma$ ($^{\rm o}$)  \\
\hline
$R\bar{3}$ & 7.317 & 7.317 & 7.094  & 120.00    \\
\hline
\multicolumn{5}{c}{Atomic positions} \\
Atoms & Site & $x$  & $y$ & $z$ \\
\hline
K &  3b & 0.00000 & 0.00000 & 0.50000 \\
P &  3a & 0.00000 & 0.00000 & 0.00000 \\
F & 18f & 0.79089 & 0.89837 & 0.13364 \\
\end{tabular}
\end{ruledtabular}
\label{TableS4}
\end{table}